\documentclass[aps,prb,reprint,10pt,superscriptaddress,showpacs]{revtex4-1}

\usepackage{graphicx}
\usepackage{amsmath}
\usepackage{amssymb}
\usepackage{color}
\usepackage[normalem]{ulem} 
\usepackage{cancel}

\newcommand{\ket}[1]{\left|#1\right>}

\begin{document}

\title{Maximizing the purity of a qubit evolving in an anisotropic environment}

\author{Xiaoya Judy Wang}
\affiliation{Department of Physics, McGill University, Montreal, Quebec, H3A 2T8, Canada}
\author{Stefano Chesi}
\affiliation{Beijing Computational Science Research Center, Beijing 100084, China}
\affiliation{CEMS, RIKEN, Wako, Saitama 351-0198, Japan}
\affiliation{Department of Physics, McGill University, Montreal, 
Quebec, H3A 2T8, Canada}
\author{W.~A.~Coish}
\affiliation{Department of Physics, McGill University, Montreal, Quebec, H3A 2T8, Canada}
\affiliation{Quantum Information Science Program, Canadian Institute for Advanced Research, Toronto, Ontario, M5G 1Z8, Canada}

\date{\today}

\begin{abstract}
We provide a general method to calculate and maximize the purity of a qubit interacting with an anisotropic non-Markovian environment. Counter to intuition, we find that the purity is often maximized by preparing and storing the qubit in a \emph{superposition} of non-interacting eigenstates. For a model relevant to decoherence of a heavy-hole spin qubit in a quantum dot or for a singlet-triplet qubit for two electrons in a double quantum dot, we show that preparation of the qubit in its non-interacting ground state can actually be the \emph{worst} choice to maximize purity. We further give analytical results for spin-echo envelope modulations of arbitrary spin components of a hole spin in a quantum dot, going beyond a standard secular approximation. We account for general dynamics in the presence of a pure-dephasing process and identify a crossover timescale at which it is again advantageous to initialize the qubit in the non-interacting ground state. Finally, we consider a general two-axis dynamical decoupling sequence and determine initial conditions that maximize purity, minimizing leakage to the environment. 
\end{abstract}

\pacs{03.65.Yz,76.60.Lz,73.21.La}

\maketitle

\section{Introduction}

A source of high-quality pure ancilla qubits is an essential element in a wide variety of applications in quantum information science. Pure ancillas are required to introduce redundancy into quantum error-correcting codes,\cite{Calderbank1996,Steane1996,Knill1997,Chiaverini2004,Fowler2009,Fowler2012} for the preparation of Greenberger-Horne-Zeilinger (GHZ) states for quantum-enhanced precision measurements, \cite{Huelga1997,Giovannetti2006} as a low-entropy resource for algorithmic cooling,\cite{Boykin2002,Baugh2005,Simmons2011} and to perform high-fidelity qubit readout.\cite{Schaetz2005,Robledo2011,Pla2013,Danjou2014} 

Despite the importance of having high-quality ancillas, it is often taken for granted that high-purity ancillas can be prepared by allowing a physical qubit system to fall into its non-interacting ground state in contact with a thermal bath at low temperature. For this reason, the preparation of an ancilla in the computational basis is often assumed to be easy relative to the more difficult task of preserving the coherence of an arbitrary qubit state. However, qubits that couple strongly to a complex environment can become correlated with the environment in a way that significantly reduces purity due to leakage to environmental degrees of freedom. Qubits that are manipulated on a time scale that is short compared to a typical thermal equilibration time may not even reach equilibrium. Although there now exist methods to mitigate the effects of somewhat impure ancilla qubits in quantum error correction schemes,\cite{Criger2012} for all of the applications stated above, it is important to prepare and store ancilla qubits in a way that maximizes their purity.

When a qubit and its environment are initialized in a factorized pure initial state, a reduction in the purity of the qubit characterizes entanglement between the qubit and its environment. In this case, the purity can be used as a measure of non-classical correlations that develop during the evolution of the qubit with its environment and hence can distinguish truly quantum from classical dynamics. This topic has become especially interesting in the context of spin-bath dynamics.\cite{Schliemann2002,Coish2007,Stanek2013,Fink2014} For slowly-evolving nuclear-spin baths, it is indeed possible to approach pure-state initial conditions through algorithmic cooling\cite{Boykin2002,Baugh2005,Simmons2011} or direct measurement of the bath state,\cite{Coish2004,Klauser2006,Giedke2006,Stepanenko2006,Robledo2011,Shulman2014} so studying the purity for these systems is especially important from both a practical and a fundamental point of view. 

\begin{figure}
\includegraphics[width = 0.45\textwidth]{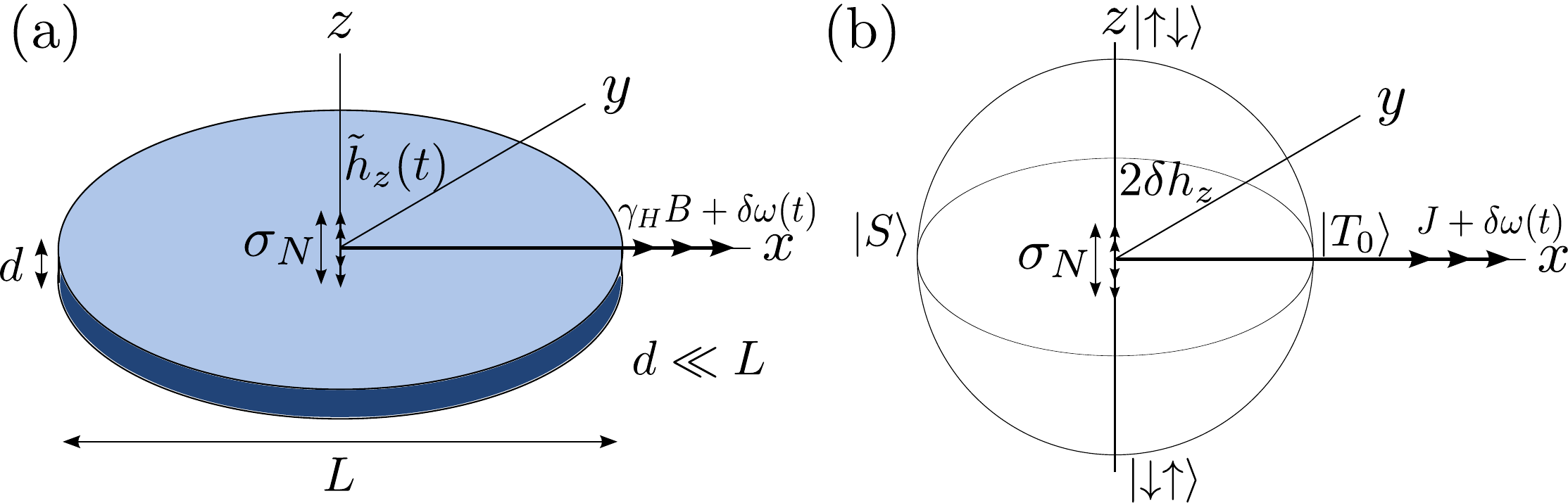}
\caption{(Color online) (a) A hole spin in a flat unstrained quantum dot having thickness $d$ much smaller than width $L$, subjected to a magnetic field of magnitude $B$ applied in-plane and a hyperfine-induced nuclear Overhauser field $h_z$ fluctuating with amplitude $\sigma_N$. (b) Bloch sphere for a singlet-triplet ($S$-$T_0$) qubit subject to a fluctuating nuclear difference field $\delta h_z$ and exchange coupling $J$ with fluctuations $\delta\omega(t)$ due to charge noise.\cite{Coish2005} All results for the $S$-$T_0$ model follow directly from the hole-spin model with the replacements $h_z\to 2\delta h_z$, $\gamma_H B\to J$, $\gamma_j\to 0$. }
\label{fig:geometry}
\end{figure}

As we show below, the evolution of qubit purity becomes highly nontrivial for a qubit interacting with a slow anisotropic environment.  Anisotropic hyperfine couplings between a central qubit spin and environmental spins are important for nitrogen-vacancy centers in diamond,\cite{Childress2006,Felton2009,Zhang2011} electrons bound to phosphorus donor impurities in silicon,\cite{Ivey1975,Witzel2007} electrons in graphene or carbon nanotubes,\cite{Yazyev2008,Fischer2009} and especially for hole spins in III-V semiconductors or silicon.\cite{Fischer2008,Fischer2010,Wang2012,Chekhovich2012,Chesi2013} Heavy-hole spins can indeed approach the extreme-anisotropic limit of a pure Ising-like coupling to nuclear spins.\cite{Fischer2008} Finally, singlet-triplet ($S$-$T_0$) qubits, describing two electrons in a double quantum dot, are described by precisely the same anisotropic decoherence model\cite{Coish2005} as a heavy-hole spin qubit (see Fig.~\ref{fig:geometry}).  

Coherence properties of single hole spins in quantum dots have been probed in detail only relatively recently.\cite{Chesi2013} Measurements of a coherent-population-trapping dip\cite{Brunner2009,Houel2014} have suggested long hole-spin coherence times, $\gtrsim 100\,\mathrm{ns}$. These measurements have been supported by time-domain studies for single-hole spin echoes\cite{DeGreve2011,Carter2014} and mode-locking or spin-echo measurements for ensembles.\cite{Fras2012,Varwig2013} Alternative measurements of hole-spin dynamics have been performed through spin-noise spectroscopy, revealing a probable anisotropic decay of hole-spin coherence.\cite{Li2012,Dahbashi2012}

In addition to optical coherent control of hole spins in self-assembled quantum dots,\cite{Brunner2009,DeGreve2011,Greilich2011,Godden2012} there are several suggestions for electrical manipulation of hole spins.\cite{Bulaev2007,Budich2012,Szumniak2012} Such electrical control has recently been demonstrated for hole spins in III-V nanowire quantum dots,\cite{Pribiag2013} and coherence times have now been measured for hole spins in Ge-Si core-shell nanowire quantum dots.\cite{Higginbotham2014} The very recent achievement of the few-hole regime in lateral gated double-dot devices,\cite{Tracy2014} suggests that previous highly successful measurements performed for electron spins\cite{Petta2005,Koppens2006,Pioro2008,Bluhm2010,Brunner2011,Shulman2012} can now be performed for hole spins, which show promise for much longer coherence times.\cite{Fischer2008,Wang2012} 

In the rest of this paper, we introduce a general method that can be used to calculate and enhance the purity of a qubit interacting with an anisotropic environment. We apply this method to the experimentally relevant problems of heavy-hole and singlet-triplet ($S$-$T_0$) spin-echo and dynamical-decoupling dynamics. Counter to common intuition, we find that preparation of the hole spin in its Zeeman ground state can be the \emph{worst} choice if the goal is to maximize purity. This surprising result is not limited to the problem of hole-spin echoes. On quite general grounds, the ideal choice to maximize purity will typically not be initialization in the eigenbasis of the isolated qubit Hamiltonian at sufficiently short time, and whenever pure-dephasing processes are weak or absent.

The rest of this paper is organized as follows. In Sec.~\ref{sec:SpinPurity} we review properties of the purity and derive general conditions to achieve the maximum purity at time $t$, starting with a factorized initial state of the qubit and a generic environment. In Sec.~\ref{sec:Born-Markov} we illustrate the method in the limit of a Born-Markov approximation, leading to exponentially decaying correlations. We demonstrate that, even in this limit, it can be suboptimal to store a qubit in the non-interacting eigenbasis. In Sec.~\ref{sec:model} we calculate spin-echo dynamics for a non-Markovian model relevant to either a heavy-hole spin in a quantum dot or a singlet-triplet ($S$-$T_0$) qubit formed by two electrons in a double quantum dot (see Fig.~\ref{fig:geometry}). In Sec.~\ref{sec:purity} we give general conditions to maximize qubit purity in a spin-echo experiment. In Sec.~\ref{sec:dephasing}, we consider the more general case of purity/coherence decay accounting for a pure-dephasing process in addition to anisotropic hyperfine coupling. In Sec.~\ref{sec:DynamicalDecoupling} we generalize the approach to a two-axis dynamical decoupling sequence and illustrate the method on the same model valid for heavy-hole or $S$-$T_0$ qubits. We conclude in Sec.~\ref{sec:Conclusion} with a summary of the main results. Technical details are given in Appendices \ref{sec:approx_gneq0}-\ref{sec:DDFilterFunctions}.

\section{Qubit Purity}\label{sec:SpinPurity}
Here we give a brief introduction to the key observable that we will evaluate, the qubit purity, $P(t)$. In addition to its importance for the preparation of high-quality ancillas in quantum error correction schemes,\cite{Bravyi2005} purity characterizes the ability to extract a finite qubit polarization after interacting with a bath for a time $t$. Maximizing the purity is essential for any scheme that aims to maximize the storage-and-retrieval fidelity of a qubit interacting with an uncontrolled environment. 

The purity of a two-level system (a qubit) is generally defined as\cite{Prosen2002}
\begin{equation}
P(t) = \mathrm{Tr}\left[\rho_S^2(t) \right] = \frac{1}{2} + 2|\langle \mathbf{S}(t) \rangle |^2.
\label{eq:purity}
\end{equation}
Here, $\rho_S(t) = \frac{1}{2}\sigma_0 + \langle \mathbf{S}(t) \rangle \cdot \boldsymbol\sigma$ is the reduced density matrix of a qubit, where $\sigma_0$ is the identity and $\boldsymbol\sigma$ is the vector of Pauli matrices. For a pure state of the qubit, the Bloch vector $\left<\mathbf{S}(t)\right>$ lies on the surface of the Bloch sphere, $\left|\left<\mathbf{S}(t)\right>\right|=1/2$, giving $P(t)=1$, while a mixed state has $\left|\left<\mathbf{S}(t)\right>\right|<1/2$ giving $P(t)<1$. Understanding the dynamics of the length of the Bloch vector, $\left|\left<\mathbf{S}(t)\right>\right|$, therefore allows for a direct evaluation of the purity, $P(t)$. In particular, we can establish a set of criteria that maximize the purity to avoid information loss.

Provided both a qubit and its environment are initially prepared in a pure state, entanglement between the qubit and environment can be characterized by the von Neumann (entanglement) entropy,
\begin{eqnarray}
E[\rho_S] &=& -\mathrm{Tr}\rho_S\log_2\rho_S=-\sum_{s=\pm}p_s\log_2 p_s,\label{eq:Entanglement}\\
p_\pm &=& \frac{1}{2}\left(1\pm 2\left|\left<\mathbf{S}(t)\right>\right|\right). 
\end{eqnarray} 
Here, $p_\pm $ give the eigenvalues of $\rho_S$. In this case, a reduction in the length of the Bloch vector $\left|\left<\mathbf{S}(t)\right>\right|<1/2$ (equivalently, $P(t)<1$) characterizes a finite degree of entanglement, $E[\rho_S]\ne 0$.\cite{Schliemann2002} Provided the environment itself can be prepared in a pure state, the purity $P(t)$ is therefore also an important measure of non-classical evolution. 

We assume the dynamics of $\left<\mathbf{S}(t)\right>$ are generated by a Hamiltonian
\begin{equation}
H(t)=H_0(t)+V(t),\quad H_0(t)=H_S(t)+H_E, 
\end{equation} 
where here, the Hamiltonians $H_S(t)$ and $H_E$ act only on the system and environment Hilbert spaces, respectively, and the perturbation $V(t)$ typically couples the two spaces. $H_S(t)$ and $V(t)$ are generally time-dependent to account for control pulses and classical noise, but we will assume $H_S(t)$ commutes with itself for all times, $\left[H_S(t),H_S(t')\right]=0$. For factorized initial conditions $\rho(0)=\rho_S(0)\otimes\rho_E(0)$ with initial system (environment) density matrix $\rho_{S(E)}(0)$, we can write generally
\begin{equation}\label{eq:SpinEvolution}
\left<\mathbf{S}(t)\right> = \left<\left[e^{-\frac{1}{2}\mathcal{L}(t)}\tilde{\mathbf{S}}(t)\right]\right>_S. 
\end{equation}
Here, the interaction picture is defined (setting $\hbar=1$) by 
\begin{equation}
\tilde{O}(t)=U_0(t)OU_0^\dagger(t),\quad U_0(t)=e^{i\int_0^t dt'H_0(t')}, 
\end{equation}
corresponding to an SO(3) rotation matrix $\left[R_0(t)\right]$ applied to the vector $\mathbf{S}=\left(S_x,S_y,S_z\right)^T$:
\begin{equation}\label{eq:InteractionPictureR0}
\tilde{\mathbf{S}}(t) = U_0(t)\mathbf{S} U_0^\dagger(t) = \left[R_0(t)\right]\cdot\mathbf{S}.
\end{equation}
We use the notation 
\begin{equation}
\left<\cdots\right>_{S(E)}=\mathrm{Tr}_{S(E)}\left[\rho_{S(E)}(0)\cdots\right]
\end{equation} 
for an average over the initial state of the system (environment). The time evolution is generated by a superoperator that acts exclusively on the qubit space:
\begin{equation}\label{eq:LDefinition}
e^{-\frac{1}{2}\mathcal{L}(t)} = \left<\mathcal{T}e^{i\int_0^t dt' L_V(t')}\right>_E.
\end{equation}
$\mathcal{T}$ is the usual time-ordering operator. The interaction-picture Liouvillian $L_V(t)$ is defined by its action on an arbitrary operator $O$ through
\begin{equation}
L_V(t)O=\left[\tilde{V}(t),O\right].
\end{equation} 

The action of Eq.~\eqref{eq:LDefinition} can generally be described by an affine map (see Ref.~\onlinecite{Nielsen2000})\footnote{The result in Ref.~\onlinecite{Nielsen2000} is written in the form (adapted to the notation used here), $\left[O\right]\cdot\left[S\right]$, where $\left[O\right]$ is a rotation matrix and $\left[S\right]$ is a real symmetric matrix. Since a real symmetric matrix can always be diagonalized by an orthogonal transformation, $\left[S\right] = \left[R^{-1}\right]\left[M\right]\left[R\right]$, we find the term on the right-hand side of Eq.~\eqref{eq:AffineMap} is $\left[O\right]\cdot\left[S\right]=\left[R'\right]\cdot\left[M\right]\cdot\left[R\right]$ with $\left[R'\right]=\left[O\right]\cdot\left[R^{-1}\right]$}
\begin{equation}\label{eq:AffineMap}
e^{-\frac{1}{2}\mathcal{L}(t)}\mathbf{S} = \left[R'(t)\right]\cdot\left[M(t)\right]\cdot\left[R(t)\right]\cdot\mathbf{S} +\overline{\left<\delta\mathbf{S}(t)\right>}.
\end{equation}
Here, $\left[R(t)\right]$ and $\left[R'(t)\right]$ are SO(3) rotation matrices. $\left[M(t)\right]$ is a magnification matrix that is diagonal with real eigenvalues,
\begin{equation}\label{eq:MDefinition}
\left[M(t)\right] = \begin{pmatrix}
e^{-\lambda_1(t)/2} & 0 & 0\\
0 & e^{-\lambda_2(t)/2} & 0\\
0 & 0 & e^{-\lambda_3(t)/2}
\end{pmatrix}.
\end{equation}
The inhomogeneous term in Eq.~\eqref{eq:AffineMap}, $\overline{\left<\delta\mathbf{S}(t)\right>}$, typically sets the long-time equilibrium value of the spin, $\overline{\left<\delta\mathbf{S}(t\to\infty)\right>}$, which is independent of the initial state for an ergodic system. For systems interacting with a sufficiently high-temperature thermal environment, the inhomogeneous term may be negligible, 
\begin{equation}\label{eq:InhomogeneousZero}
\overline{\left<\delta\mathbf{S}(t)\right>}\simeq 0.
\end{equation} 
Indeed, this turns out to be the case in the experimentally relevant problems of a hole-spin or $S$-$T_0$ qubit interacting with an unpolarized nuclear-spin bath, which we address below (see also Appendix \ref{sec:approx_gneq0}). Thus, we first proceed under the (realistic) assumption that Eq.~\eqref{eq:InhomogeneousZero} is satisfied. 

Inserting Eq.~\eqref{eq:InteractionPictureR0} into Eq.~\eqref{eq:SpinEvolution}, and applying Eq.~\eqref{eq:AffineMap} with $\overline{\left<\delta\mathbf{S}(t)\right>}=0$ gives 
\begin{equation}\label{eq:SMatrices}
\left<\mathbf{S}(t)\right>=\left[R_0(t)\right]\cdot\left[R^\prime(t)\right]\cdot\left[M(t)\right]\cdot\left[R(t)\right]\cdot\left<\mathbf{S}(0)\right>.
\end{equation}
The first two rotations, $\left[R_0(t)\right]\cdot\left[R^\prime(t)\right]$, preserve the length of the Bloch vector, so they will not enter into the formula for purity. This leaves
\begin{equation}\label{eq:Ssqr}
\left|\left<\mathbf{S}(t)\right>\right|^2=\sum_\mu \left[M(t)\right]^2_{\mu\mu}\left(\left[R(t)\right]\cdot\left<\mathbf{S}(0)\right>\right)_\mu^2.
\end{equation}
The effect of $\left[R(t)\right]$ is to align the Bloch vector along principal axes defined by a set of mutually orthogonal unit vectors $\mathbf{\hat{e}}_\mu (t)$ (see, e.g., Fig.~\ref{fig:eigens}, below), 
\begin{equation}\label{eq:e-mu-definition}
\mathbf{\hat{e}}_\mu(t)\cdot\mathbf{S} = \left(\left[R(t)\right]\cdot\mathbf{S}\right)_\mu.
\end{equation}
Using Eqs.~\eqref{eq:MDefinition}, \eqref{eq:Ssqr}, and \eqref{eq:e-mu-definition} in Eq.~\eqref{eq:purity} then gives a compact form for the purity,
\begin{equation}\label{eq:PurityEigenvalueDecomposition}
P(t) = \frac{1}{2}+2\sum_\mu e^{-\lambda_\mu(t)}\left|\left<\mathbf{S}(0)\right>\cdot\mathbf{\hat{e}}_\mu(t)\right|^2.
\end{equation}
The purity of the qubit at time $t$ therefore depends on the eigenvalues $\lambda_\mu(t)$ and on the initial conditions through $\left<\mathbf{S}(0)\right>\cdot\mathbf{\hat{e}}_\mu(t)$. In particular, it is always possible to maximize $P(t)$ by choosing to initialize the qubit along a direction $\mathbf{\hat{e}}_\mu(t)$ associated with the smallest eigenvalue, $\lambda_\mu(t)<\lambda_\nu(t)\quad (\mu\ne\nu)$. In this case, the purity is given simply by
\begin{equation}\label{eq:PmuGeneral}
P_\mu (t) =\frac{1}{2}\left(1+e^{-\lambda_\mu(t)}\right),\quad \left|\left<\mathbf{S}(0)\right>\cdot\mathbf{\hat{e}}_\mu(t)\right|=1/2.
\end{equation} 

Note that the general case of finite $\overline{\left<\delta \mathbf{S}(t)\right>}$ is not significantly more complex---in this case, we simply need to find the initial state $\left<\mathbf{S}(0)\right>$ that maximizes the magnitude
\begin{equation}
\left|\left<\mathbf{S}(t)\right>\right|=\left|\left[R'(t)\right]\cdot\left[M(t)\right]\cdot\left[R(t)\right]\cdot\left<\mathbf{S}(0)\right>+\overline{\left<\delta \mathbf{S}(t)\right>}\right|.
\end{equation}
However, the result for this general case cannot be expressed in the simple form of Eq.~\eqref{eq:PurityEigenvalueDecomposition}.

When $\mathcal{L}(t)$ can be expressed as a real symmetric matrix ($\left[\mathcal{L}\right]_{\alpha\beta}=\left[\mathcal{L}\right]_{\beta\alpha}$, where $\left[\mathcal{L}\right]_{\alpha\beta}=2\mathrm{Tr}\left\{S_\alpha\mathcal{L}S_\beta\right\}$), this matrix is diagonalized with an orthogonal rotation, i.e., $\left[R'(t)\right]=\left[R^{-1}(t)\right]$ in Eq.~\eqref{eq:AffineMap}. In this case, the parameters $\lambda_\mu(t)$ are the real eigenvalues of the superoperator $\mathcal{L}(t)$ and the unit vectors $\mathbf{\hat{e}}_\mu(t)$ determine the associated eigenoperators through
\begin{equation}\label{eq:LEigenvalue}
\mathcal{L}(t)\left[\mathbf{\hat{e}}_\mu (t)\cdot \mathbf{S}\right] =\lambda_\mu(t)\left[\mathbf{\hat{e}}_\mu (t)\cdot \mathbf{S}\right].
\end{equation}
Decomposing the spin operator $\mathbf{S}$ in terms of its components along the unit vectors $\mathbf{\hat{e}}_\mu$ then gives a simplified expression for the spin expectation values when $\left[\mathcal{L}\right]$ is symmetric,
\begin{equation}\label{eq:spinexpect}
\left<\mathbf{S}(t)\right>=\sum_\mu e^{-\lambda_\mu(t)/2}\mathbf{\hat{e}}_\mu (t)\cdot \left<\mathbf{S}(0)\right> \left[R_0(t)\right]\cdot\mathbf{\hat{e}}_\mu (t).
\end{equation}
The case of a real symmetric generator $\mathcal{L}(t)$ will be relevant to the example of hole-spin or $S$-$T_0$ qubit dynamics, which we address in the following sections.

To solve the eigenvalue equation, Eq.~\eqref{eq:LEigenvalue}, it is first necessary to derive a suitable approximation for the superoperator $\mathcal{L}(t)$, defined by Eq.~\eqref{eq:LDefinition}. As will be shown below, when $H_0(t)$ generates sufficiently rapid oscillations in $\tilde{V}(t)$, a leading-order Magnus expansion can be performed on the time-ordered exponential in Eq.~\eqref{eq:LDefinition}. For a sufficiently large environment with initial state described by many uncorrelated degrees of freedom, the moments associated with the average $\left<\cdots\right>_E$ will be approximately Gaussian. When $\left<L_V(t)\right>_E=0$, the combination of these two approximations leads to 
\begin{equation}\label{eq:LApprox}
\mathcal{L}(t) \simeq \mathcal{L}_0(t)= \int_0^tdt_1\int_0^tdt_2\left<L_V(t_1)L_V(t_2)\right>_E.
\end{equation}
Of course, the applicability of Eq.~\eqref{eq:LApprox} depends sensitively on the details of the physical system under study. In the remaining sections we will evaluate and justify this formula for a model with a pure Ising-like anisotropic hyperfine coupling. This model is directly relevant to hole spins in quantum dots or to singlet-triplet ($S$-$T_0$) qubits formed by two electrons in double quantum dots (see Fig.~\ref{fig:geometry}).

Higher-order terms in the Magnus expansion involve progressively more integrals over the oscillating perturbation $\tilde{V}(t)$. The leading-order Magnus expansion can therefore always be justified at sufficiently short time. A general sufficient condition for convergence of the Magnus expansion is\cite{Blanes2009}
\begin{equation}
\int_0^t dt' ||\tilde{V}(t')||_2 <\pi.
\end{equation}  
Convergence of the expansion is then generally guaranteed for $t<t_\mathrm{max}$ where
\begin{equation}\label{eq:tmax-general}
t_\mathrm{max} = \frac{\pi}{\mathrm{max}\left[||\tilde{V}(t')||_2\right]},
\end{equation}
and where $\mathrm{max}\left[||\tilde{V}(t')||_2\right]$ is the maximum of $||\tilde{V}(t')||_2$ on the interval $[0,t]$. In practice, Eq.~\eqref{eq:tmax-general} often drastically underestimates the range of applicability of the leading-order Magnus expansion. When all terms in $\tilde{V}(t)$ are rapidly oscillating about zero with typical amplitude $\delta\omega_\mathrm{rms}$ and typical fast frequency $\omega$, a direct analysis of the higher-order terms leads to the condition\cite{Wang2012}
\begin{equation}\label{eq:tmaxDefinition}
t\lesssim \tau_\mathrm{max}= \frac{\omega}{\delta\omega_\mathrm{rms}^2}.
\end{equation}
The fast frequency $\omega$ may be given by the precession frequency for a spin system. Alternatively, in the case of a dynamical decoupling sequence (which we consider in Sec.~\ref{sec:DynamicalDecoupling}), the fast frequency may be given by $\omega\sim 1/\tau$, where $\tau$ is the time between decoupling pulses. We have found the analysis leading to Eq.~\eqref{eq:tmaxDefinition} to accurately reflect the time scale of failure of the Magnus expansion for, e.g., the free-induction decay of a hole spin in a quantum dot.\cite{Wang2013}

\section{Born-Markov limit} \label{sec:Born-Markov}

Equation \eqref{eq:LApprox} generally accounts for nonstationary and non-Markovian dynamics. This is necessary for a slow environment and a qubit subjected to a dynamical decoupling sequence. Before considering this more general scenario, here we explore the consequences of a Born-Markov approximation, appropriate to the limit of weak coupling to an environment with a short bath correlation time compared to the relevant dephasing ($T_\phi$) and relaxation ($T_1$) times. We assume a qubit with system Hamiltonian $H_S=\omega S_z$, so that the computational basis states, $\ket{0}=\ket{\Downarrow}$ and $\ket{1}=\ket{\Uparrow}$, are associated with the Bloch vector aligned along the $\hat{\mathbf{z}}$-axis. Bloch-Redfield theory then gives the nonvanishing superoperator matrix elements and inhomogeneous term in the interaction picture:
\begin{eqnarray}
\left[\mathcal{L}(t)\right]_{xx}=\left[\mathcal{L}(t)\right]_{yy} & \simeq & \left(\frac{1}{T_\phi}+\frac{1}{2T_1}\right)2 t,\\
\left[\mathcal{L}(t)\right]_{zz} & \simeq & \frac{2t}{T_1},\\
\overline{\left<\delta S_z(t)\right>} & \simeq & \left(1-e^{-t/T_1}\right)\overline{\left<S_z\right>},
\end{eqnarray}
with long-time steady-state value $\overline{\left< S_z\right>}$.

In this limit, a spin prepared along $+\hat{\mathbf{z}}$ will decay according to 
\begin{equation}
\left<S_z(t)\right> = (1/2-\overline{\left< S_z\right>})e^{-t/T_1}+\overline{\left< S_z\right>}.
\end{equation}
For a spin prepared along $+\hat{\mathbf{x}}$, and in the absence of pure dephasing ($1/T_\phi = 0$): 
\begin{equation}
\left<S_x(t)\right> = (1/2)e^{-t/2T_1}. 
\end{equation} 
Expanding for short time, we find $\left|\left<S_x(t)\right>\right|>\left|\left<S_z(t)\right>\right|$ at leading order in $t/T_1$ whenever $\overline{\left< S_z\right>}<1/4$. Thus, in the absence of a pure-dephasing process and for a sufficiently high-temperature environment (so that $\overline{\left<S_z\right>}$ is sufficiently small), it is advantageous to prepare the system in a superposition of non-interacting eigenstates, even in the case of Markovian decay.

In the next section, we consider the more general case of a slow non-Markovian environment. For a non-Markovian system-environment evolution, purity can be lost and recovered through a series of revivals that can be induced through a dynamical-decoupling sequence. The optimization over initial conditions in this case is therefore nontrivial, but practically useful.

\section{Non-Markovian dynamics: Heavy-hole and $S$-$T_0$ spin echo} \label{sec:model}

Here we will present a model of non-Markovian dynamics, allowing the recovery of qubit purity at later times through recurrences. We will primarily focus on the dynamics of a heavy-hole spin in a quantum dot. However, a special limit of the model is directly relevant to $S$-$T_0$ qubits formed by two electrons in a double quantum dot.

\subsection{Hole-spin model}\label{sec:Hole-spin-model}

We consider a model Hamiltonian, appropriate for a heavy-hole spin qubit in a flat semiconductor quantum dot\cite{Wang2012} with an in-plane magnetic field applied along the $x$-axis [see Fig.~\ref{fig:geometry}(a)]:

\begin{align}
\label{eq:fullH}
H &= H_\mathrm{Z}+ H_\mathrm{hf},\\
H_\mathrm{Z} &= -\gamma_H B S_x - \sum_{j,k_j} \gamma_{j} B I_{k_j}^x,\\
H_\mathrm{hf} &= h_zS_z = \sum_{j,k_j} A_{k_j} I_{k_j}^z S_z.
\end{align}
In the sums above, $j$ labels the distinct nuclear isotopes and $k_j$ labels the set of sites occupied by nuclear spins of isotope $j$. Here, $H_\mathrm{Z}$ gives the hole-spin and nuclear-spin Zeeman terms, and $H_\mathrm{hf}$ describes the anisotropic hyperfine interaction between hole and nuclear spins. 
$\mathbf{S}=\boldsymbol{\sigma}/2$ is a pseudospin-$1/2$ operator in the heavy-hole subspace, and $\mathbf{I}_k$ is the nuclear spin at site $k$. The hole gyromagnetic ratio is $\gamma_H = \mu_B g_\bot$, with $\mu_B$ the Bohr magneton and $g_\bot$ the in-plane hole $g$-factor. The gyromagnetic ratio of the nucleus at site $k_j$ of isotopic species $j$ having total spin $I_j$ is denoted by $\gamma_{j}$. The hyperfine couplings, $A_{k_j}$, are given by $A_{k_j} = A^{j} v_0 |\psi(\mathbf{r}_{k_j})|^2$, where $A^j$ is the hyperfine coupling for nuclear species $j$, $v_0$ is the volume per nuclear spin, and $\psi(\mathbf{r}_k)$ is the heavy-hole envelope wavefunction evaluated at site $\mathbf{r}_k$. For a Gaussian envelope function in two dimensions,\cite{Coish2004}
\begin{equation}
A_k \simeq \frac{A}{N} e^{-k/N}, ~~ k = 0,1,2,.,
\end{equation}
where $N$ is the number of nuclear spins within a quantum-dot Bohr radius. 

\begin{figure}
\includegraphics[width = 0.48\textwidth]{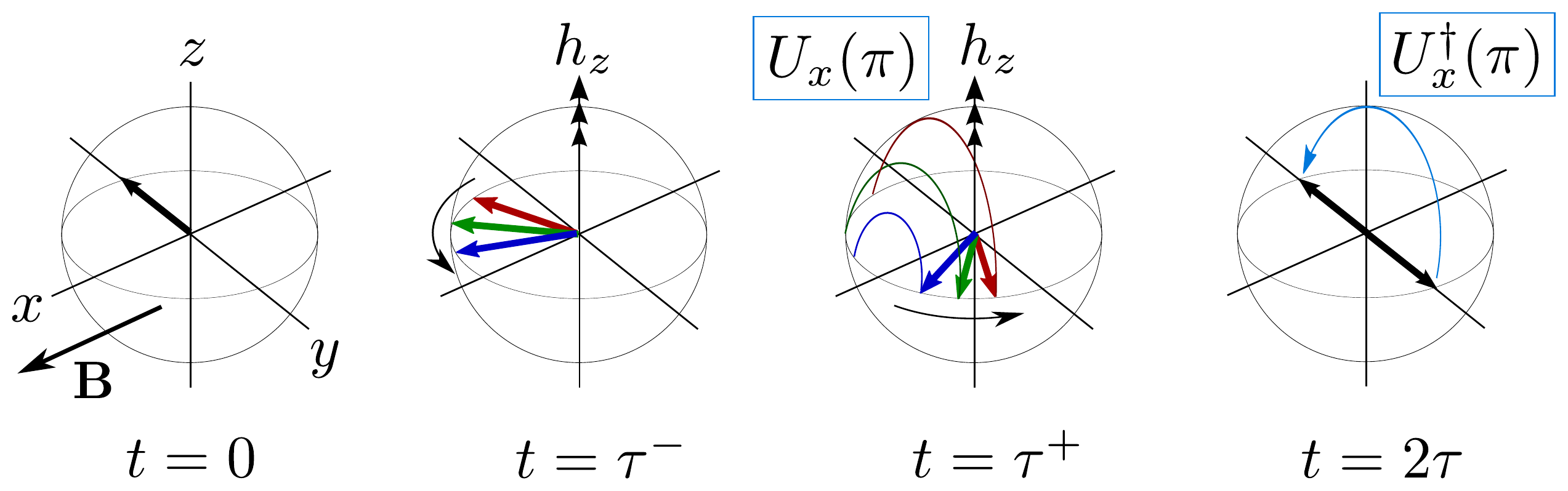}
\caption{(Color online) Hahn echo sequence. $\pi$-rotations about $\mathbf{\hat{x}}$ reverse dephasing from static fluctuations in $h_z$ along $\mathbf{\hat{z}}$ due to the hyperfine interaction. For the sake of clarity we have assumed $g_\bot \simeq 0$ for this illustration so that there is no precession of the hole spin about $\mathbf{\hat{x}}$. 
\label{fig:echo}}
\end{figure}

Further assuming a uniform distribution of different nuclear species across the dot and $N \gg 1$ (typically $N \simeq 10^4 - 10^6$), we define the average hyperfine constant $A$ as
\begin{equation}
A = \sum_k A_k \simeq \sum_j \nu_j A^j,
\end{equation}
where $\nu_j$ is the isotopic abundance of species $j$. In numerical estimates we will assume, for simplicity, a single average value $A$ of the hyperfine constant corresponding to $\nu_j$ for an $\mathrm{In_{0.5}Ga_{0.5}As}$ quantum dot\cite{Fischer2008} ($A \simeq 13\mu\mathrm{eV}$), and $\gamma_j,\,I_j$ appropriate for natural abundances of isotopes of Ga, As, and In taken from Table 1 of Ref.~\onlinecite{Coish2009}. For heavy holes, the ratio $|A/A^{(e)}|$ of hole to electron hyperfine coupling strengths has been estimated theoretically \cite{Fischer2008} in GaAs and confirmed experimentally \cite{Chekhovich2011a,Fallahi2010} in InGaAs and InP/GaInP to be of order $|A/A^{(e)}| \sim 0.1$. This is consistent with $A \simeq 13\,\mu e\mathrm{V}$ since $A^{(e)}\simeq\,90\,\mu e\mathrm{V}$ in GaAs.\cite{Paget1977}

Random fluctuations in the nuclear field cause rapid hole-spin decoherence via the hyperfine coupling described above. A spin-echo sequence can remove fluctuations that are approximately static over the time scale of hole-spin preparation and measurement. A Hahn echo sequence corresponds to a free evolution for time $t<\tau$, application of a $\pi$-rotation about the $x$-axis, $U_x(\pi)$, at $t=\tau$, followed by another free evolution for time $t\in\left(\tau,2\tau\right)$. We consider a second $\pi$-rotation, $U_x^\dagger(\pi)$, at $t=2\tau$ to return the spin to its original orientation (see Fig.~\ref{fig:echo}). Noting that $U_x(\pi)S_zU_x^\dagger(\pi)=-S_z$, but $U_x(\pi)S_xU_x^\dagger(\pi)=S_x$, we account for the Hahn echo sequence illustrated in Fig.~\ref{fig:echo} with the identifications:
\begin{equation}\label{eq:HoleSpinH0andV}
H_0 = H_Z,\quad V(t) = s(t)H_\mathrm{hf},
\end{equation}
where
\begin{align}
s(t) &= \begin{cases}
+1 &~~~~ 0\leq t < \tau,\\
-1 &~~~~ \tau\leq t \leq 2\tau. 
\label{eq:echo-s}
\end{cases}
\end{align}
With the associations given in Eq.~\eqref{eq:HoleSpinH0andV}, we can now apply the analysis of Sec.~\ref{sec:SpinPurity} to the problem of Hahn echo, using the leading-order Magnus expansion and Gaussian approximation to obtain the approximate generator $\mathcal{L}(2\tau)\simeq\mathcal{L}_0(2\tau)$ given in Eq.~\eqref{eq:LApprox}. See Refs.~\onlinecite{Wang2012,Beaudoin2013,Chesi2013} for further details on implementing the Magnus expansion and Gaussian approximation specific to this problem.

To make analytical progress, we rewrite the superoperator $\mathcal{L}_0$ in matrix form. In the basis of spin-1/2 operators, $\{S_x, S_y, S_z\}$, the matrix elements $\left[\mathcal{L}_0(2\tau)\right]_{\alpha\beta}$ are given by
\begin{equation}
\mathcal{L}_0(2\tau) S_\alpha = \sum_\beta \left[\mathcal{L}_0(2\tau)\right]_{\beta\alpha} S_\beta; ~~ \alpha, \beta = x,y,z.
\label{eq:LMatrixElements}
\end{equation}
The matrix $\left[\mathcal{L}_0(2\tau)\right]$ can be found explicitly in terms of bath correlation functions $\left<\mathcal{B}_\alpha(2\tau)\mathcal{B}_\beta(2\tau)\right>$, with bath operators $\mathcal{B}_\alpha$ defined by (see Appendix \ref{sec:approx_gneq0}):
\begin{equation}\label{eq:BathOperatorsDefinition}
\int_0^{2\tau}dt\tilde{V}(t)=\sum_\alpha \mathcal{B}_\alpha(2\tau) S_\alpha.
\end{equation}
We assume the initial state of the nuclear-spin bath describes uncorrelated spins without second-order coherences and with vanishing polarization, so that
\begin{eqnarray}
\left<h_j^+h_{j'}^+\right>&=&\left<h_j^-h_{j'}^-\right>=0,\label{eq:no-correlations}\\
\left<h_j^+h_{j'}^-\right>&=&\left<h_j^-h_{j'}^+\right>=2\sigma_j^2\delta_{jj'}.\label{eq:sigma}
\end{eqnarray}
Here we have introduced the nuclear field operators over an isotope $j$,
\begin{equation}
\mathbf{h}_j = \sum_{k_j} A_{k_j}\mathbf{I}_{k_j},\quad h_j^\pm = h_j^y\pm i h_j^z.
\end{equation}
Equation \eqref{eq:sigma} above defines the nuclear-field fluctuation $\sigma_j$ due to isotope $j$. For the purposes of studying system-bath entanglement, it may be interesting to prepare a \emph{pure} state of the bath and observe the resulting purity dynamics [see the discussion leading to Eq.~\eqref{eq:Entanglement} above]. We note that the conditions given in Eqs. \eqref{eq:no-correlations} and \eqref{eq:sigma} will be approximately satisfied for a pure state with suitably random initialization (e.g., by choosing a random orientation for each nuclear spin independently). For practical measurements, the initial conditions of the nuclear-spin bath are often well-described by an infinite-temperature thermal state, for which
\begin{equation}
\sigma_j^2 = \frac{I_j(I_j+1)}{3}\sum_{k_j}(A_{k_j})^2. 
\end{equation}
For explicit estimates, we will make use of the total nuclear-field variance,
\begin{equation}
\sigma_N^2 = \sum_j \sigma_j^2.
\end{equation}

\subsection{Mapping to an $S$-$T_0$ qubit}\label{sec:SingletTripletMapping}

As illustrated in Fig.~\ref{fig:geometry}, the model presented here for heavy-hole spin dynamics and decoherence can be mapped exactly onto a well-studied model of singlet-triplet decoherence.\cite{Coish2005} In particular, the heavy-hole spin-$S_z$ eigenstates $\left|\Uparrow\right>$ and $\left|\Downarrow\right>$ can be associated with two-electron states $\left|\uparrow\downarrow\right>$ and $\left|\downarrow\uparrow\right>$ for two electron spins in a double quantum dot, making up the singlet $\ket{S}$ and triplet $\ket{T_0}$ states:
\begin{eqnarray}
\ket{\Uparrow}&\to & \ket{\uparrow\downarrow},\\
\ket{\Downarrow}&\to & \ket{\downarrow\uparrow},\\
\ket{S}&=&\frac{1}{\sqrt{2}}\left(\ket{\uparrow\downarrow}-\ket{\downarrow\uparrow}\right),\\
\ket{T_0}&=&\frac{1}{\sqrt{2}}\left(\ket{\uparrow\downarrow}+\ket{\downarrow\uparrow}\right).
\end{eqnarray} 

The following associations for energy scales complete the mapping:
\begin{eqnarray}
\gamma_H B\to J,\\
h_z\to 2\delta h_z,\\
\gamma_j\to 0.
\end{eqnarray}
Here, $J$ is the exchange coupling, $\delta h_z$ is the nuclear difference field between the two quantum dots, and for a double quantum dot subject to a uniform magnetic field, $\delta h_z$ commutes with the nuclear-spin Zeeman term, leading to $\gamma_j=0$. Inhomogeneities in the magnetic field in this case could lead to dynamics in $\delta h_z$, which can then act back on the $S$-$T_0$ qubit. This effect has been investigated recently in Ref.~\onlinecite{Beaudoin2013}, but we neglect it here for simplicity.

\subsection{Hahn-echo dynamics}\label{sec:SpinEchoDynamics}
As suggested by the form of Eq.~\eqref{eq:BathOperatorsDefinition}, the bath operators can be conveniently rewritten in terms of the complex-valued filter functions
\begin{eqnarray}
Z_{j\pm}(2\tau) & = & \sigma_j\int_0^{2\tau} dt s(t) e^{i\omega_{j\pm} t}\label{eq:ZiDefinition}\\
& = & -i\frac{4\sigma_j}{\omega_{j\pm}}\sin^2\frac{\omega_{j\pm}\tau}{2}e^{i\omega_{j\pm}\tau},\label{eq:ZiHahn}\\
\omega_{j\pm} & = & (\gamma_H\pm\gamma_j)B.
\end{eqnarray}
In the conventional theory of spin-echo decay, functions such as Eq.~\eqref{eq:ZiDefinition} determine a \emph{filter function} $\mathcal{F}$, that restricts the frequency-content of the noise that can act to dephase a qubit through the absolute magnitude of $Z_{j\pm}$: \cite{uhrig2007,deSousa2009,Cywinski2008,Bylander2011,Alvarez2011}
\begin{equation}
\mathcal{F}(\omega_{j\pm},2\tau) \propto |Z_{j\pm}(2\tau)|^2.
\end{equation}
Here, we will find both the magnitude and the phase of the functions $Z_{j\pm}(2\tau)$ will be essential in determining spin dynamics. While the magnitude of the functions $Z_{j\pm}(2\tau)$ will modify the spectral content of the noise, the phase of these functions will be crucial in determining a set of principal axes that determine the anisotropy of the decay process. Quite significantly, we will be able to exploit information about this decay anisotropy to identify optimal initialization/storage protocols to maximize the purity of a spin qubit.

Direct evaluation of the bath correlation functions and application of the relationships derived in Appendix \ref{sec:approx_gneq0} gives the matrix
\begin{equation}
\left[\mathcal{L}_0(2\tau)\right] = \begin{pmatrix}
\lambda_x(2\tau) & 0 \\[4pt]
0 & \left[\mathcal{L}^{yz}(2\tau)\right]
\end{pmatrix}.
\label{eq:liouvillian}
\end{equation}
Here, one eigenvalue of the superoperator is $\lambda_x(2\tau)$. The $2\times 2$ submatrix $\left[\mathcal{L}^{yz}(2\tau)\right]$ can be written as 
\begin{equation}\label{eq:LyzDefinition}
\left[\mathcal{L}^{yz}(2\tau)\right] = \frac{1}{2}\left[\lambda_x(2\tau) \tau_0 + \mathrm{Re}Z^2(2\tau) \tau_3 + \mathrm{Im}Z^2(2\tau) \tau_1\right],
\end{equation}
where we have introduced the $2\times 2$ identity matrix $\tau_0$ and usual Pauli matrices $\tau_\mu $. In Eq.~\eqref{eq:LyzDefinition}, we have also introduced the complex function $Z(2\tau)$:
\begin{equation}\label{eq:ZDefinition}
Z^2(2\tau) = \sum_j Z_{j+}(2\tau)Z_{j-}(2\tau).
\end{equation}

\begin{figure}
\includegraphics[width = 0.3\textwidth]{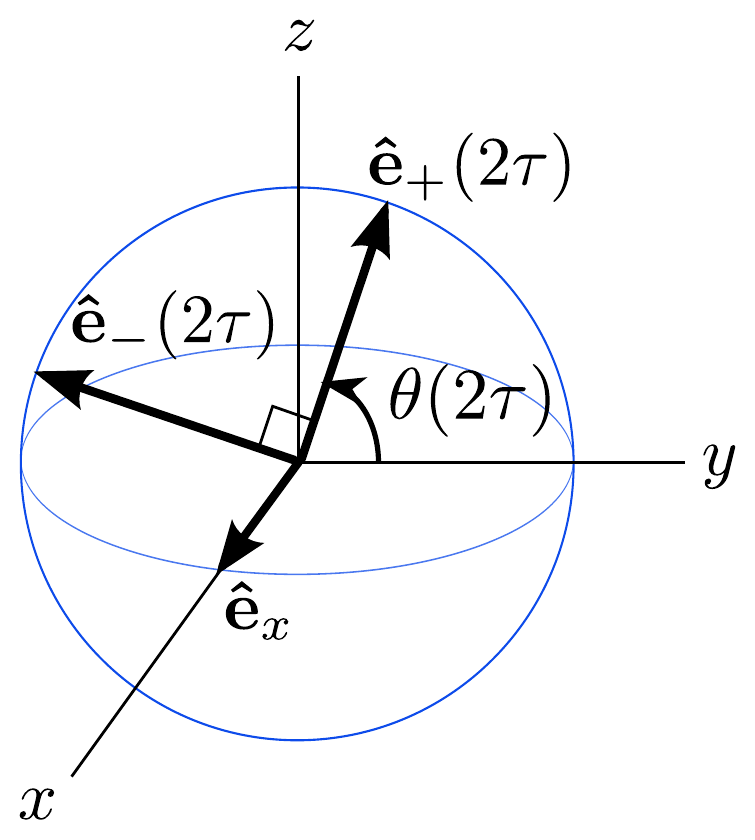}
\caption{(Color online) Unit vectors satisfying the eigenvalue equation, Eq.~\eqref{eq:eigeneq}, forming an orthonormal basis. $\mathbf{\hat{e}}_x=\mathbf{\hat{x}}$ while $\mathbf{\hat{e}}_+(2\tau)$ and $\mathbf{\hat{e}}_-(2\tau)$ correspond to $\mathbf{\hat{y}}$ and $\mathbf{\hat{z}}$ rotated by an angle $\theta(2\tau)$ [given by Eq.~(\ref{eq:theta})] about the $x$-axis.}
\label{fig:eigens}
\end{figure}

By diagonalizing the matrix in Eq.~\eqref{eq:liouvillian}, we solve Eq.~\eqref{eq:LEigenvalue} for the eigenvalues, $\lambda_\mu(2\tau)$, and vectors, $\mathbf{\hat{e}_\mu}(2\tau)$, with $\mu=x,\pm$:
\begin{equation}
\mathcal{L}_0(2\tau) \left[\mathbf{S \cdot \hat{e}_\mu}(2\tau)\right] = \lambda_\mu(2\tau) \left[\mathbf{S \cdot \hat{e}_\mu}(2\tau)\right].
\label{eq:eigeneq}
\end{equation}
We find the eigenvalues: 
\begin{eqnarray}
\lambda_x(2\tau) &=& \frac{1}{2}\sum_j \left(|Z_{j+}(2\tau)|^2+|Z_{j-}(2\tau)|^2\right),\label{eq:lambdax}\\
\lambda_\pm(2\tau) &=& \frac{1}{2}\left(\lambda_x(2\tau)\pm \left|Z(2\tau)\right|^2\right).\label{eq:lambdapm-general}
\end{eqnarray}
When either $\gamma_H>\gamma_j$ or $\gamma_j>\gamma_H$ for all nuclear-spin species $j$, Eq. \eqref{eq:lambdapm-general} becomes
\begin{equation}
\lambda_\pm(2\tau) = \frac{1}{4}\sum_j\left(|Z_{j+}(2\tau)|\pm|Z_{j-}(2\tau)|\right)^2.\label{eq:lambdapm}
\end{equation}
The associated unit vectors (illustrated in Fig.~\ref{fig:eigens}) are
\begin{eqnarray}
\mathbf{\hat{e}}_x &=& \mathbf{\hat{x}},\label{eq:ex}\\
\mathbf{\hat{e}}_+(2\tau) &=& \cos \theta(2\tau)\mathbf{\hat{y}} + \sin \theta(2\tau)\mathbf{\hat{z}},\label{eq:eplus}\\
\mathbf{\hat{e}}_-(2\tau) &=& -\sin \theta(2\tau)\mathbf{\hat{y}} + \cos \theta(2\tau)\mathbf{\hat{z}}.\label{eq:eminus}
\end{eqnarray} 
The angle $\theta(2\tau)$ is determined by the SU(2) rotation that diagonalizes Eq.~\eqref{eq:LyzDefinition}. This angle is given by
\begin{equation}\label{eq:theta}
\theta(2\tau) = \arg Z(2\tau).
\end{equation}
The vectors $\mathbf{\hat{e}}_\mu (2\tau)$ are parametrized by the time $\tau$ between $\pi$-pulses. However, we stress that these are not dynamical quantities, evolving during the echo sequence. Instead, $\mathbf{\hat{e}}_\mu (2\tau)$ determines the \emph{initial condition} for a spin that should be chosen to recover a given purity $P_\mu (2\tau)$ after a Hahn echo sequence.

Inserting Eq.~\eqref{eq:ZiHahn} for the complex-valued filter functions $Z_{j\pm}(2\tau)$ into Eq.~\eqref{eq:ZDefinition} shows that $\theta(2\tau)$ can generally alternate between two simple forms:
\begin{align}\label{eq:theta-two-vals}
\theta(2\tau) = \left\{
\begin{array}{l}
\gamma_H B\tau-\frac{\pi}{2},\quad \mathcal{G}(2\tau)>0\\
\gamma_H B\tau,\quad \mathcal{G}(2\tau)<0
\end{array}\right.,
\end{align}
with
\begin{equation}
\mathcal{G}(2\tau) = \sum_j\frac{\sigma_j^2\sin^2\left(\frac{\omega_{j+}\tau}{2}\right)\sin^2\left(\frac{\omega_{j-}\tau}{2}\right)}{\left(\gamma_H^2-\gamma_j^2\right)B^2}.
\end{equation}

Alignment of the spin along $\mathbf{\hat{e}}_-(2\tau)$ at time $t=0$ will maximize the purity at time $t=2\tau$ since $\lambda_-(2\tau)$ gives the smallest eigenvalue [see Eqs. \eqref{eq:lambdax}, \eqref{eq:lambdapm-general}]. Referring to Fig. \ref{fig:eigens} and Eq. \eqref{eq:theta-two-vals} for the angle $\theta(2\tau)$, we see that when the nuclear-spin system can be taken as approximately static compared to the hole-spin precession, $\gamma_H>\gamma_j$ (giving $\mathcal{G}>0$), a spin initialized along $\mathbf{\hat{e}}_-(2\tau)$ will advance in time at an angle $\phi = \theta(2\tau)+\pi/2-\gamma_H B t= \gamma_H B(\tau-t)$ from the $y$-axis. The spin will then be aligned with the $y$-axis at the time of the first $\pi$-pulse ($t=\tau$). That this choice is optimal can be simply understood from a semiclassical model of a fluctuating magnetic field along $z$ in the limit of a purely static nuclear field and arises from the perfect symmetry of this problem for reflections through the $x$-axis within the $x$-$y$ plane (see Appendix \ref{sec:SimpleExample}). This scenario ($\gamma_j\simeq 0$) applies exactly to the case of an $S$-$T_0$ qubit in a uniform magnetic field (see Sec.~\ref{sec:SingletTripletMapping}). In the opposite limit of a slow hole-spin precession compared to the nuclear-spin precession, $\gamma_H<\gamma_j$ ($\mathcal{G}<0$), the optimal choice is to prepare the hole spin so that it aligns with the $z$-axis at the first $\pi$-pulse. This result is easy to understand in the limit $\gamma_H=0$, since in this case the $z$-component of hole spin is a constant of the motion, and is therefore preserved for all time. More generally, when $\gamma_H\sim\gamma_j$, the optimal initialization axis will alternate nontrivially as a function of $\tau$ to favor alignment with either $\mathbf{\hat{y}}$ or $\mathbf{\hat{z}}$ at the time of the first $\pi$-pulse. When we consider additional pure-dephasing processes in Section \ref{sec:dephasing} below, we will find such a nontrivial behavior even when $\gamma_H\gg\gamma_j$, the limit typically realized in current experiments.

For this problem, spin dynamics in the $S_x$-subspace have been discussed previously.\cite{Wang2012} In this subspace, we find
\begin{equation}
\label{eq:SxL0}
\frac{\langle S_x(2\tau) \rangle} {\langle S_x(0) \rangle} \simeq e^{-\frac{1}{2}\lambda_x(2\tau)}.
\end{equation}
A motional-averaging regime is reached for $\lambda_x\lesssim 1$, corresponding to $\omega\gtrsim \sigma_N $, where $\sigma_N\sim A/\sqrt{N}$ is the typical amplitude of nuclear-field fluctuations and $\omega = B\cdot\mathrm{max}\{\gamma_i,\gamma_H\}$ gives the frequency of rapid oscillations. In this regime, the hole spin experiences envelope modulations\footnote{These modulations have a similar origin to electron spin-echo envelope modulation (ESEEM), well-known in the spin-resonance literature. However, the effect here is distinct from the usual approach to ESEEM since we have accounted for leading non-secular corrections through the leading-order Magnus expansion. Without these corrections, the modulation would vanish for the highly anisotropic interaction considered here.} with amplitude $\sim \lambda_x\sim |\sigma_N/\omega|^2<1$. From Eq.~\eqref{eq:SxL0}, it is already clear that a hole spin initially aligned along $\mathbf{\hat{x}}$ will have a purity that is modulated in time according to the envelope modulations. 

Further setting $g_\bot\mu_\mathrm{B}=\gamma_H=0$ in the expressions above, Eq.~(\ref{eq:SxL0}) recovers the result previously given in Ref.~\onlinecite{Wang2012},
\begin{equation}\label{eq:magnus}
\frac{\langle S_x(2\tau) \rangle} {\langle S_x(0) \rangle}\simeq
\exp{\left[ -\sum_j \frac{8\sigma_j^2}{(\gamma_j B)^2} \sin^4{\left(\frac{\gamma_j B \tau}{2}\right)}\right]}.
\end{equation}

We now evaluate $\left<S_y(2\tau)\right>$ and $\left<S_z(2\tau)\right>$ by inverting Eqs.~\eqref{eq:eplus} and \eqref{eq:eminus} for $\mathbf{\hat{e}}_+(2\tau)$ and $\mathbf{\hat{e}}_-(2\tau)$,
\begin{align}
S_y &= \cos\theta(2\tau)\mathbf{S\cdot \hat{e}_+}(2\tau) -\sin\theta(2\tau)\mathbf{S\cdot \hat{e}_-}(2\tau), \\
S_z &= \sin\theta(2\tau)\mathbf{S\cdot \hat{e}_+}(2\tau) +\cos\theta(2\tau)\mathbf{S\cdot \hat{e}_-}(2\tau).
\end{align}
The evolution takes a simple form in terms of the operators $S'_\pm$:
\begin{equation}
S'_\pm = S_y \pm iS_z.
\end{equation}
The spin evolution in the $y$-$z$ plane is then described by
\begin{equation}\label{eq:SPrime}
\langle S'_+(2\tau) \rangle = e^{i\phi(2\tau)} \sum_{\mu=\pm} \sqrt{\mu} e^{-\frac{1}{2}\lambda_\mu(2\tau) }\langle \mathbf{S}(0) \rangle \cdot \mathbf{\hat{e}}_\mu(2\tau),
\end{equation}
where $\phi(2\tau) = -\gamma_H B 2\tau + \theta(2\tau)$, $\sqrt{+} = \sqrt{1} = 1$, and $\sqrt{-} = \sqrt{-1} = i$. The phase $\phi(2\tau)$ tracks the mismatch in evolution of the interaction-picture rotating frame and the rotation to principal axes for the generator $\mathcal{L}_0(2\tau)$ (see Fig.~\ref{fig:eigens}). The eigenvalues $\lambda_\mu(2\tau)$ control the degree of damping/modulation in the amplitude of the spin. 

The spin dynamics under the action of the anisotropic interactions presented here are strongly dependent on the initial direction of the spin and on the measurement axis. We will find it convenient to parametrize the initial state for a spin in the $y$-$z$ plane by an angle $\varphi$ between the $y$-axis and the initial spin vector:
\begin{equation}
\left<S_+^\prime(0)\right>=\left<S_y(0)\right>+i\left<S_z(0)\right>=\frac{1}{2}e^{i\varphi}.
\end{equation} 
For a spin prepared at an angle $\varphi$ to the $y$-axis, we define the coherence factor in the rotating frame,
\begin{equation}
C_{\varphi}(2\tau)= 2 e^{i\gamma_H B2\tau}\left<S_+^\prime (2\tau)\right>,\quad\left<S_+^\prime(0)\right>=\frac{1}{2}e^{i\varphi}. 
\end{equation}
We then find the general expression for this coherence factor,
\begin{multline}\label{eq:Cphi0}
C_{\varphi}(2\tau) = e^{-\lambda_x(2\tau)/4+i\varphi}\left[\cosh\left(\frac{|Z(2\tau)|^2}{4}\right)-\right.\\
\left.-e^{-i2\left[\varphi-\theta(2\tau)\right]}\sinh\left(\frac{|Z(2\tau)|^2}{4}\right)\right].
\end{multline}
The first term in Eq.~\eqref{eq:Cphi0} [$\propto \cosh\left(|Z|^2/4\right)$] varies slowly in the rotating frame, while the second term [$\propto \sinh\left(|Z|^2/4\right)$] experiences violent modulations at a frequency determined by the hole-spin Zeeman energy, $2\theta(2\tau)\sim \gamma_H B2\tau$, due to non-secular ``counter-rotating" corrections. This second contribution, $\sim |Z|^2\ll 1$, evolves slowly in the lab frame, in spite of the hole-spin Zeeman term. Thus, while the first term would likely decay rapidly due to electric-field-induced fluctuations in the hole Zeeman energy, as reported in recent experiments,\cite{DeGreve2011,Houel2014,Carter2014} we expect some contribution from the second term to survive this dephasing mechanism. Such a pure-dephasing process is investigated in detail in Sec.~\ref{sec:dephasing} below.

Equation \eqref{eq:Cphi0} recovers the expected results for initialization along one of the principal axes: $\varphi=\varphi_\pm(2\tau)$, corresponding to alignment of the initial spin with $\mathbf{\hat{e}}_\pm (2\tau)$ (see Fig.~\ref{fig:eigens}):
\begin{eqnarray}
C_{\varphi_\pm} (2\tau) & = & \exp\left[i\varphi_\pm(2\tau)-\lambda_{\pm}(2\tau)/2\right],\\
\varphi_+(2\tau) &=& \theta(2\tau),\\
\varphi_-(2\tau) &=& \theta(2\tau) +\pi/2.
\end{eqnarray}

\begin{figure}
\includegraphics[width = 0.45\textwidth]{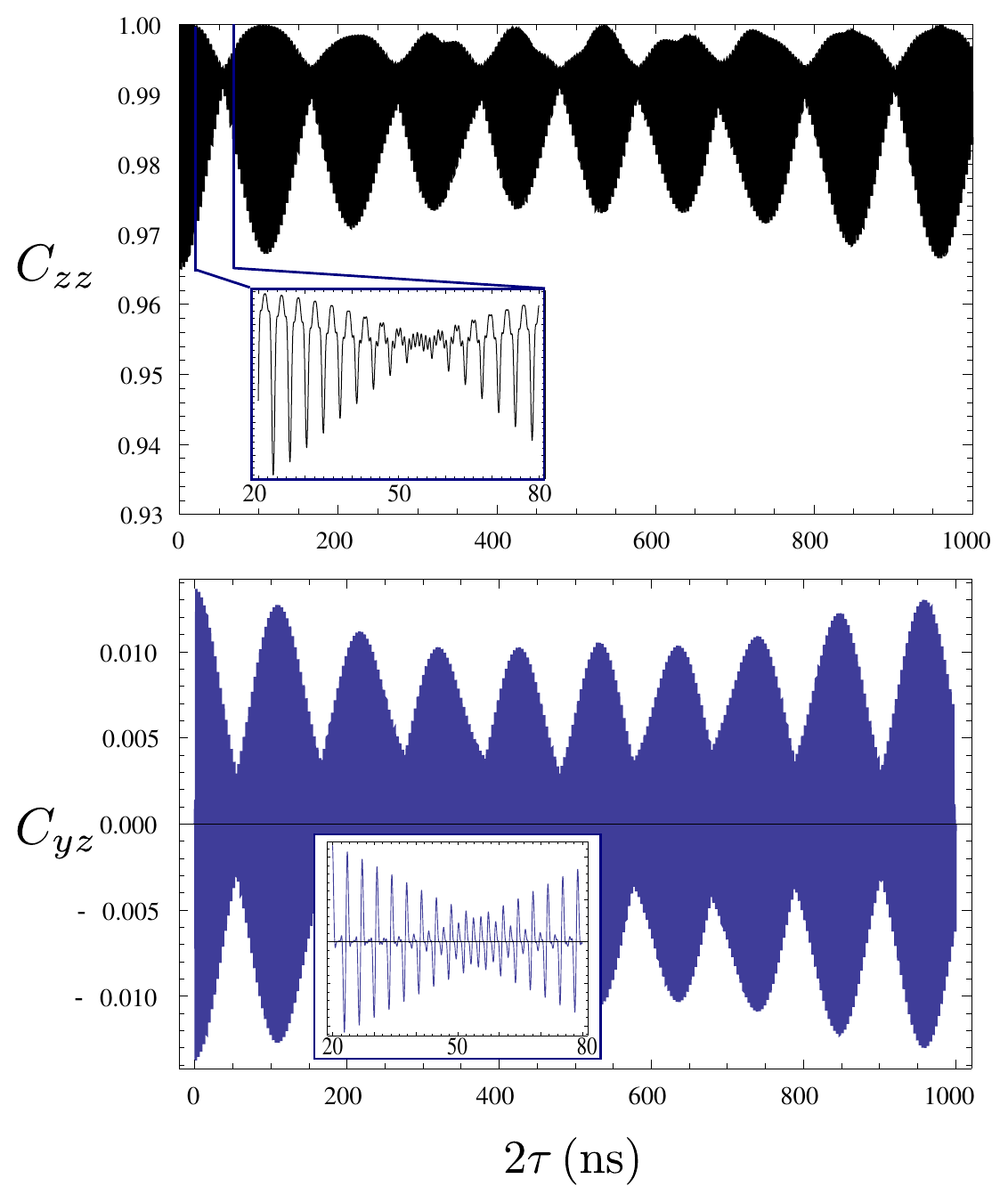}
\caption{(Color online) Correlators in the rotating frame [Eqs.~\eqref{eq:Cyy}-\eqref{eq:Cyz}], assuming an in-plane magnetic field $B=1\,\mathrm{T}$, with in-plane hole-spin g-factor $g_\bot = 0.04$, for an $\mathrm{In}_x\mathrm{Ga}_{1-x}\mathrm{As}$ quantum dot containing $N = 10^4$ nuclear spins assuming uniform In doping $x=0.5$, and nuclear gyromagnetic ratios $\gamma_j$ and total nuclear spins $I_j$ from Ref.~\onlinecite{Coish2009} appropriate for this material.}
\label{fig:correlators-rotating}
\end{figure}

To explore the general spin dynamics of this problem, in which no special care has been taken to initialize the spin along one of the principal axes $\mathbf{\hat{e}}_\pm$, we define the general correlators $C_{\alpha\beta}$ corresponding to the coherence for initialization along axis $\beta$ and measurement along axis $\alpha$ in the rotating frame:
\begin{eqnarray}
C_{yy}(2\tau) &=& \mathrm{Re}\left[C_{\varphi=0}(2\tau)\right],\label{eq:Cyy}\\
C_{zy}(2\tau) &=& \mathrm{Im}\left[C_{\varphi=0}(2\tau)\right],\label{eq:Czy}\\
C_{zz}(2\tau) &=& \mathrm{Im}\left[C_{\varphi=\pi/2}(2\tau)\right],\label{eq:Czz}\\
C_{yz}(2\tau) &=& \mathrm{Re}\left[C_{\varphi=\pi/2}(2\tau)\right].\label{eq:Cyz}
\end{eqnarray}
Correlators such as those given above have been measured, for example, in recent experiments on hole spins in single quantum dots.\cite{Carter2014} Those experiments showed similar modulations as seen here, although the authors of Ref.~\onlinecite{Carter2014} have interpreted the modulations in their data in terms of a dynamic nuclear polarization effect. Two of the correlators above are shown for typical experimental parameters in Fig.~\ref{fig:correlators-rotating}. We note that the general correlators will contain contributions from each of the eigenvalues $\lambda_\pm (2\tau)$. While each of these experiences modulations at the nuclear Larmor frequency, the modulations for $\lambda_\pm = \sum_j (|Z_{j+}|\pm|Z_{j-}|)^2/4$ are $\pi$ out of phase with respect to each other [see, e.g., the modulations of $P_+$ (determined by $\lambda_+$) relative to those for $P_-$ (determined by $\lambda_-$) in Fig.~\ref{fig:spinpurity}]. These out-of-phase modulations generally lead to a sequence of maxima at \emph{twice} the nuclear Larmor frequency, similar to the result seen for modulations in the experiment of Ref.~\onlinecite{Carter2014}. We note that the same modulations with the same frequency are predicted within this model for free-induction decay [the limit $n=0$ of an $n$-pulse dynamical decoupling sequence, see Eq.~\eqref{eq:PDDXFilter} in Appendix \ref{sec:DDFilterFunctions}]. The amplitude of modulations ($\propto 1/B^2$) is strongly suppressed in a large magnetic field $B$, so for high-field experiments, it may be difficult to see this effect. However, for $B\lesssim 1\,\mathrm{T}$ and for typical quantum-dot parameters, the modulations can be a substantial fraction of the decay, as we show here. 

\section{Maximizing purity} \label{sec:purity}

\begin{figure}
\includegraphics[width = 0.45\textwidth]{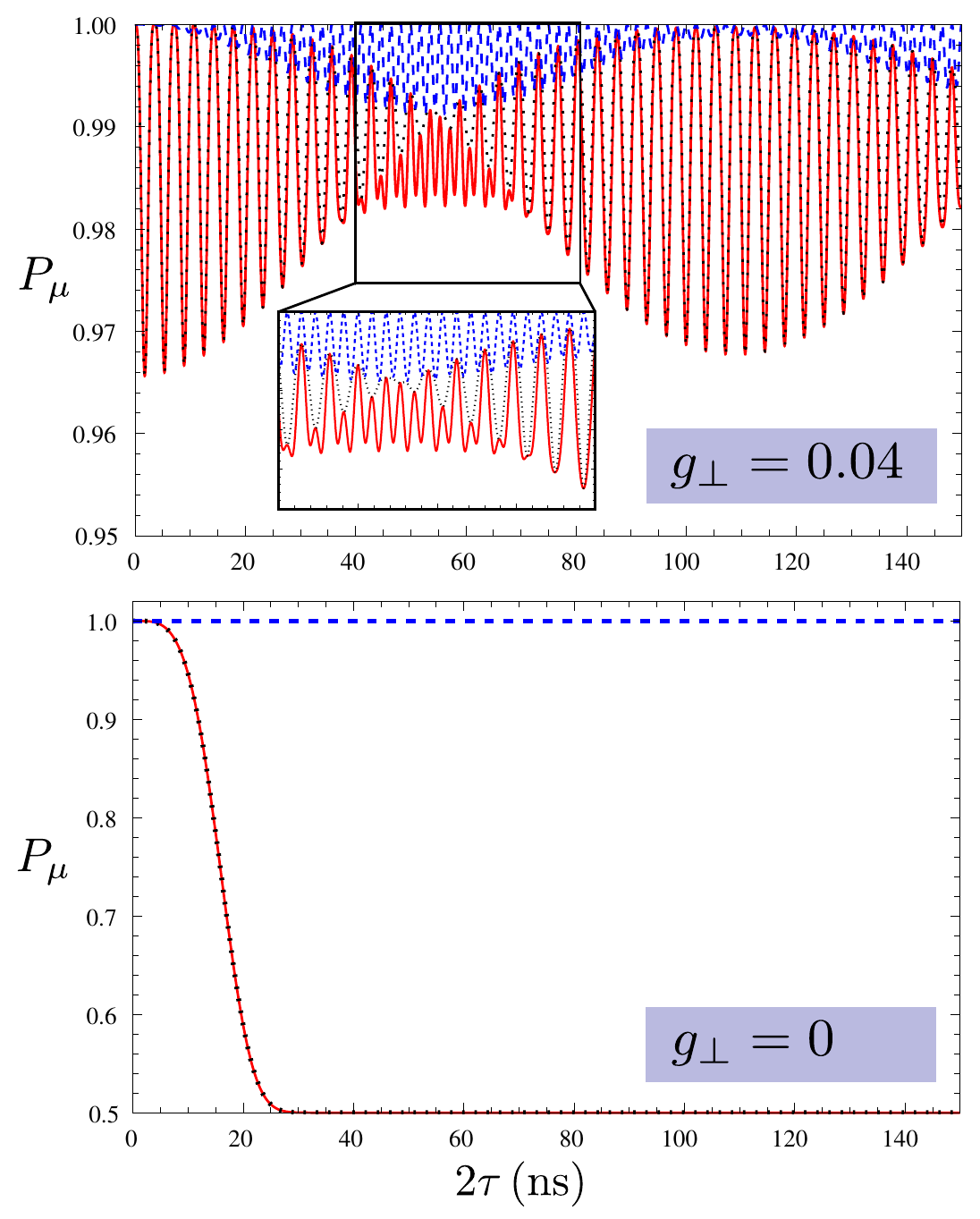}
\caption{(Color online) Spin-echo purity $P_\mu(2\tau)$ from Eq.~\eqref{eq:PmuDefinition} assuming initialization along $\mathbf{\hat{e}}_-(2\tau)$ [$P_-(2\tau)$, blue dashed line], $\mathbf{\hat{e}}_+(2\tau)$ [$P_+(2\tau)$, black dotted line], and $\mathbf{\hat{e}}_x(2\tau)=\mathbf{\hat{x}}$ [$P_x(2\tau)$, red solid line], with $B = 1\,\mathrm{T}$, $g_\bot = 0.04$, $N = 10^4$, and $\gamma_j$ and $I_j$ from Ref.~\onlinecite{Coish2009}. The purity at time $2\tau$ is maximized when initializing along $\mathbf{\hat{e}}_-(2\tau)$. Top panel: purity with a finite in-plane hole $g$-factor, $g_\bot = 0.04$. Bottom panel: purity with a vanishing in-plane hole $g$-factor, $g_\bot = 0$.}
\label{fig:spinpurity}
\end{figure}
 
The spin-echo purity, $P(2\tau)$, characterizes our ability to recover a pure ancilla qubit at a time $2\tau$ after preparation and application of a refocusing pulse. From Eq.~\eqref{eq:PurityEigenvalueDecomposition}, this quantity depends on the initialization of the qubit. In particular, if we initialize along one of the unit vectors $\mathbf{\hat{e}}_\mu(2\tau)$, we find the simple expression for the purity at time $2\tau$, as in Eq.~\eqref{eq:PmuGeneral}:
\begin{equation}\label{eq:PmuDefinition}
P_\mu(2\tau) = \frac{1}{2}\left(1+e^{-\lambda_\mu(2\tau)}\right).
\end{equation}
As discussed in Sec.~\ref{sec:SpinPurity} following Eq.~\eqref{eq:PurityEigenvalueDecomposition}, and as is clear from Eq.~\eqref{eq:PmuDefinition}, the purity of a qubit recovered at time $2\tau$ can be maximized by initializing along the direction $\mathbf{\hat{e}}_\mu(2\tau)$ associated with the smallest eigenvalue $\lambda_\mu(2\tau)$. 

Na\"{\i}vely, one might expect that the best strategy would be to prepare an ancilla qubit in an eigenstate (e.g., the ground state) of the unperturbed Hamiltonian, $H_0$. In the case of hole-spin qubits, this would correspond to preparing the spin along the applied magnetic field [along the $x$-axis for the geometry shown in Fig.~\ref{fig:geometry}(a)], $\left|0\right>=\left|\Uparrow_x\right>$ [where $S_x\left|\Uparrow_x\right>=+(1/2)\left|\Uparrow_x\right>$]. For $S$-$T_0$ qubits, this corresponds to initializing and storing in the singlet state $\ket{S}$ [see Fig.~\ref{fig:geometry}(b)]. However, in this case, with the generator given in Eq.~\eqref{eq:liouvillian}, we find the following general relationship, valid for all $\tau$ within the range of validity of the Gaussian approximation and leading-order Magnus expansion:
\begin{equation}\label{eq:lambdas}
\lambda_x(2\tau) \ge \lambda_+(2\tau) \ge \lambda_-(2\tau).
\end{equation}
The inequalities in Eq.~\eqref{eq:lambdas} follow directly from Eqs.~\eqref{eq:lambdax} and \eqref{eq:lambdapm-general}.

Quite generally, the purity is maximized by preparing the hole spin in the $y$-$z$ plane, in an equal superposition of Zeeman eigenstates:
\begin{equation}\label{eq:purity-hierarchy}
P_-(2\tau) \ge P_+(2\tau) \ge P_x(2\tau).
\end{equation} 
The three quantities in Eq.~\eqref{eq:purity-hierarchy} are shown in Fig.~\ref{fig:spinpurity} for typical experimental parameters, illustrating the inequality. In the limit $\gamma_H=\mu_B g_\perp=0$, this result can be intuitively understood. When $g_\perp =0$, $\left[S_z,H\right]=0$, so a spin initialized along the $z$-axis will be preserved for all time, while a spin initialized along $\mathbf{\hat{x}}$ or $\mathbf{\hat{y}}$ will decay due to fluctuations along $\mathbf{\hat{z}}$ (see the lower panel of Fig.~\ref{fig:spinpurity}; in this case, $\mathbf{\hat{e}}_-=\mathbf{\hat{z}}, \mathbf{\hat{e}}_+=\mathbf{\hat{y}}$). That this relationship [Eq.~\eqref{eq:purity-hierarchy}] continues to hold for $g_\perp\ne 0$ in any magnetic field and for all $\tau$ (within the range of validity of the approximations used here) is less trivially obvious. 

It is straightforward to extend the above analysis to the more general case of an arbitrary anisotropic hyperfine tensor (see Appendix \ref{sec:approx_gneq0}). In this case, when leading non-secular corrections are included using the leading-order Magnus expansion and Gaussian approximations, the Zeeman ground state will not generally be optimal for initialization. The procedure described here can be used to predict an optimal state in which to store an ancilla. This may be useful in other systems with anisotropic interactions, including nitrogen-vacancy (NV) centers in diamond,\cite{Robledo2011} or phosphorus donors in silicon,\cite{Pla2013} where the high-fidelity preparation of electron-spin ancillas is important for nuclear-spin readout.

The non-intuitive result given in Eq.~\eqref{eq:purity-hierarchy} presupposes the absence of additional decoherence mechanisms. A rapid pure-dephasing process would typically reduce the purity for states initialized perpendicular to the magnetic field, relative to those initialized along the magnetic field. One source of pure dephasing for hole spins arises due to electric-field-induced fluctuations in the Zeeman energy (equivalently, fluctuations in the exchange interaction for $S$-$T_0$ qubits). Such a mechanism has been identified as the predominant dephasing source for hole spins in Refs.~\onlinecite{DeGreve2011} and \onlinecite{Houel2014}. In the presence of a Markovian pure dephasing process that takes place on a time scale $T_\phi$, our conclusions remain valid in the limit $2\tau<T_\phi$ whenever the decay due to pure dephasing is small compared to the amplitude of envelope modulations, i.e. when $\gamma_H\gg\gamma_i$,
\begin{equation}\label{eq:PureDephasingTime}
2\tau \lesssim 2\tau_c = T_\phi\left(\frac{A}{\sqrt{N}\gamma_H B}\right)^2.
\end{equation}
For storage of ancilla qubits beyond the time scale indicated in Eq.~\eqref{eq:PureDephasingTime}, it will be advantageous to prepare the qubit in the Zeeman eigenbasis.

We consider the detailed role of a pure-dephasing process on the general dynamics of a hole-spin (equivalently, $S$-$T_0$) qubit and on purity decay in the next section.

\section{Pure dephasing} \label{sec:dephasing}

As mentioned above, pure-dephasing mechanisms can modify the results of our analysis for maximizing purity. In particular, a pure-dephasing mechanism due to a fluctuating Zeeman term through an electric-field-dependent g-factor has been identified as a primary source of decoherence for hole spins in recent experiments.\cite{DeGreve2011,Houel2014} It is straightforward to generalize the analysis of the previous sections to the case of a fluctuating Zeeman term with the replacement:\cite{Wang2012}
\begin{equation}\label{eq:ZeemanFluctuations}
\gamma_H B2\tau\to\gamma_H B2\tau+\phi(2\tau);\quad \phi(2\tau) = \int_0^{2\tau}dt\delta\omega(t), 
\end{equation}
with $\delta\omega(t)$ a Gaussian random variable describing a stationary white-noise process
\begin{equation}\label{eq:delta-omega}
\langle\langle\delta\omega(t)\delta\omega(t')\rangle\rangle = \frac{2}{T_\phi}\delta(t-t').
\end{equation} 
Here, double angle brackets $\langle\langle\cdots\rangle\rangle$ indicate an average over realizations of the noise $\delta\omega(t)$. The white-noise form given in Eq.~\eqref{eq:delta-omega} is a reasonable approximation for, e.g., Johnson-Nyquist noise due to nearby metallic gates.\cite{Langsjoen2012,Langsjoen2014} This assumption will break down for, e.g., colored noise due to slowly-varying charged impurities.\cite{Houel2014} It would be straightforward to extend the analysis presented here to the case of Gaussian colored noise. To emphasize the limitations of our earlier conclusions in the presence of pure dephasing, here we focus on the simplest (and often realistic) white-noise form given in Eq.~\eqref{eq:delta-omega}. 

Accounting for the modification to the Zeeman term, Eq.~\eqref{eq:ZeemanFluctuations}, the coherence factor in the rotating frame becomes
\begin{equation}
C_{\varphi}(2\tau) = 2\left(\langle\langle e^{-i\phi}e^{-\frac{1}{2}\mathcal{L}[\phi]}\rangle\rangle S_+^\prime\right>_S,\, \langle S_+^\prime(0)\rangle=\frac{1}{2}e^{i\varphi}.
\end{equation}
In the absence of hyperfine coupling, we would have $\mathcal{L}=0$, leaving a simple exponentially-decaying coherence factor,
\begin{equation}
C_{\varphi}(2\tau) = e^{i\varphi-\frac{1}{2}\langle\langle\phi^2(2\tau)\rangle\rangle}=e^{i\varphi-2\tau/T_\phi},
\end{equation}
where we have used the fact that the noise is Gaussian in the first step and the fact that it is white [Eq.~\eqref{eq:delta-omega}] in the second. While $\phi$ is Gaussian-distributed, $\mathcal{L}[\phi]$ is a highly nonlinear function of $\phi$, making a direct Gaussian average difficult. In general, we would like an expansion valid up to time scales $t\gtrsim T_\phi$ (giving $\langle\langle \phi\rangle\rangle \gtrsim 1$), so an expansion for small $\phi$, which was justified in evaluating the longitudinal spin $\left<S_x(2\tau)\right>$,\cite{Wang2012} is not generally possible for the coherence factor. Instead, here we perform a moment expansion, valid for $|\mathcal{L}|\lesssim 1$,
\begin{equation}
\langle\langle e^{-i\phi}e^{-\frac{1}{2}\mathcal{L}[\phi]}\rangle\rangle = \langle\langle e^{-i\phi}\rangle\rangle \langle e^{-\frac{1}{2}\mathcal{L}[\phi]}\rangle_\phi.
\end{equation}
Here, for an arbitrary operator $\mathcal{O}$, the average $\left<\cdots\right>_\phi$ is defined by
\begin{equation}
\langle \mathcal{O}[\phi]\rangle_\phi = \frac{\langle\langle e^{-i\phi}\mathcal{O}[\phi]\rangle\rangle}{\langle\langle e^{-i\phi}\rangle\rangle}=e^{\frac{1}{2}\langle\langle\phi^2\rangle\rangle}\langle\langle e^{-i\phi}\mathcal{O}[\phi]\rangle\rangle.
\end{equation}
At leading order in the moment expansion,
\begin{equation}
\langle e^{-\frac{1}{2}\mathcal{L}[\phi]}\rangle_\phi \simeq e^{-\frac{1}{2}\langle\mathcal{L}[\phi]\rangle_\phi}.
\end{equation}
From a leading-order Magnus expansion, we have $\mathcal{L}\simeq \mathcal{L}_0$ and in the regime of applicability of the Magnus expansion, $|\mathcal{L}_0(2\tau)|<1$ for all time, allowing us to neglect all higher moments with small corrections.

\begin{figure}
\includegraphics[width = 0.45\textwidth]{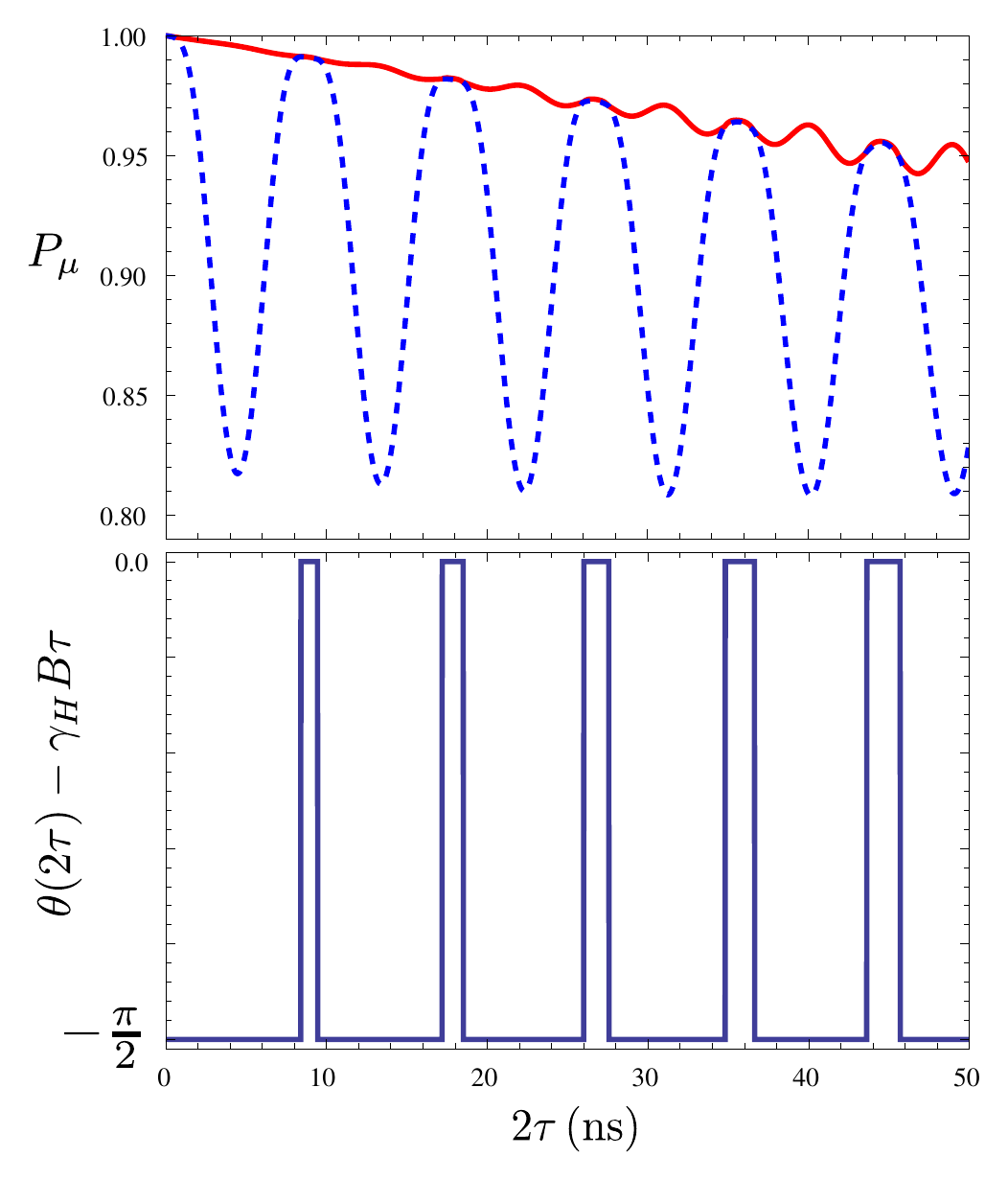}
\caption{(Color online) Top: Purities $P_-$ (red solid line) and $P_+$ (blue dashed line). Bottom: Angle $\theta(2\tau)$ determining the initial axes, as in Fig.~\ref{fig:eigens}. We have assumed a Markovian pure-dephasing process due, e.g., to electric-field noise. We have assumed a dephasing time $T_\phi=1\,\mu\mathrm{s}$ and have used the same material parameters as in Figs.~\ref{fig:correlators-rotating} and \ref{fig:spinpurity} for an $\mathrm{In_{0.5}Ga_{0.5}As}$ quantum dot but with a magnetic field of $B=400\,\mathrm{mT}$.}
\label{fig:dephasing-purities}
\end{figure}

\begin{figure}
\includegraphics[width = 0.45\textwidth]{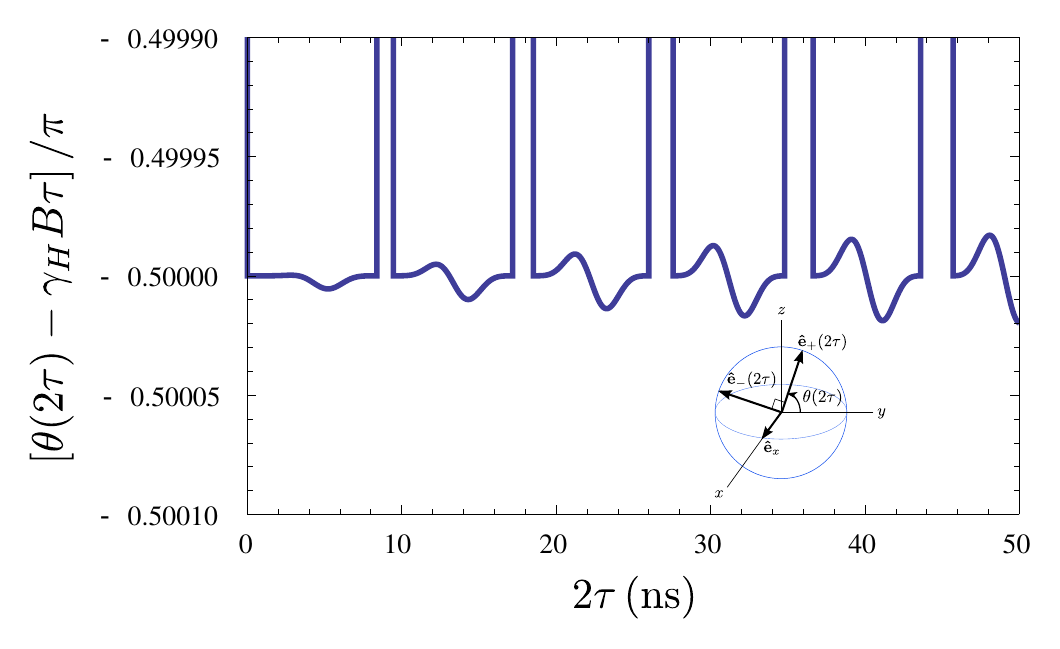}
\caption{(Color online) Detail of the angle $\theta$ for the evolution shown in Fig.~\ref{fig:dephasing-purities}. The angle $\theta$ defines the unit vectors $\mathbf{\hat{e}}_\pm$ (see inset and Fig.~\ref{fig:eigens}). Due to the presence of a pure-dephasing process, the optimal angle $\theta$ follows a complex trajectory, in general, deviating slightly from $\theta(2\tau)-\gamma_H B\tau =0,-\pi/2$.}
\label{fig:dephasing-angle-detail}
\end{figure}

Averaging over random instances of the fluctuating Zeeman term, $\delta\omega(t)$, then gives an analogous expression to Eq.~\eqref{eq:LyzDefinition}, but accounting for pure dephasing:
\begin{equation}\label{eq:Lyz0}
\langle\left[\mathcal{L}^{yz}\right]\rangle_\phi=\frac{1}{2}\left(\langle\lambda_x\rangle_\phi\tau_0+\langle\mathrm{Re}Z^2\rangle_\phi\tau_3+\langle\mathrm{Im}Z^2\rangle_\phi\tau_1\right).
\end{equation}
The averages above are evaluated explicitly in Appendix \ref{sec:AvgPureDephasing}. The coefficients in Eq.~\eqref{eq:Lyz0} are no longer real, but the matrix can nevertheless be exponentiated directly to determine the coherence factor 
\begin{multline}\label{eq:CphiDephasing}
C_{\varphi} (2\tau) = e^{-2\tau/T_\phi-\langle\lambda_x\rangle_\phi/4}\left[e^{i\varphi} \cosh\frac{Q^2}{4}-\right.\\
\left.-e^{-i\varphi}\frac{\langle Z^2\rangle_\phi}{Q^2}\sinh\frac{Q^2}{4}\right],
\end{multline}
where
\begin{equation}
Q^2 = \left[\langle\mathrm{Re} Z^2\rangle_\phi^2 +\langle\mathrm{Im} Z^2\rangle_\phi^2\right]^{1/2}.
\end{equation}
\begin{figure}
\includegraphics[width = 0.45\textwidth]{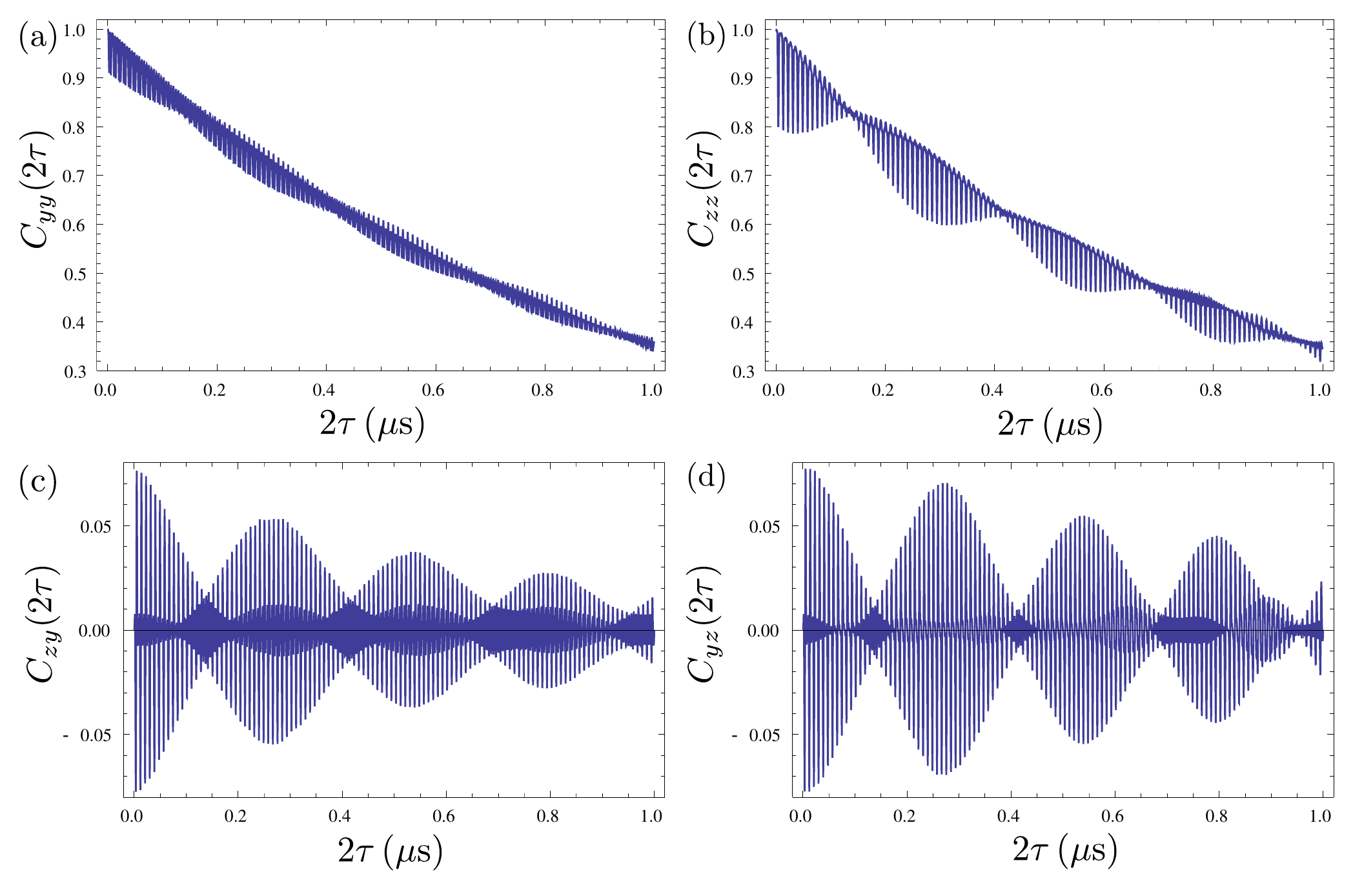}
\caption{(Color online) Correlators $C_{\alpha\beta}(2\tau)$ showing dynamics for preparation along axis $\beta$ and measurement along axis $\alpha$ in the rotating frame. The material parameters used are the same as in Figs.~\ref{fig:correlators-rotating} and \ref{fig:spinpurity} for an $\mathrm{In_{0.5}Ga_{0.5}As}$ quantum dot, but with magnetic field $B=400\,\mathrm{mT}$ and with a pure dephasing process giving rise to $T_\phi=1\,\mu\mathrm{s}$.}
\label{fig:dephasing-correlators}
\end{figure}

The angle $\theta$ is then the value $\theta=\varphi$ for which the length of the Bloch vector (set by $|C_{\varphi}|$) is minimal (corresponding to alignment along the vector $\mathbf{\hat{e}_+}$). This value can be read off directly from Eq.~\eqref{eq:CphiDephasing}, giving
\begin{equation}
\theta = \frac{1}{2}\mathrm{arg}\left[\langle Z^2\rangle_\phi\frac{\tanh Q^2/4}{Q^2}\right].
\end{equation}

The purities for a hole spin initialized in the $y$-$z$ plane are shown for typical experimental parameters in Fig.~\ref{fig:dephasing-purities} along with the angle $\theta(2\tau)$ that determines the principal axes for $\mathcal{L}_0$. In the presence of a pure-dephasing process, the optimal initialization axis alternates as a function of $\tau$ to favor alignment of the spin with either the $y$-axis at $t=\tau$ [$\theta(2\tau)-\gamma_H B\tau \simeq -\pi/2$] or the $z$-axis at $t=\tau$ [$\theta(2\tau)-\gamma_H B\tau \simeq 0$]. The additional dynamics induced through the average over random Zeeman fields gives rise to a nontrivial evolution of the angle $\theta(2\tau)$ beyond this simple picture (see Fig.~\ref{fig:dephasing-angle-detail}). While these corrections may be small here, they can be accurately determined using the procedure outlined above provided the dephasing model itself is known accurately. 

The correlators $C_{\alpha\beta}$ corresponding to initialization along direction $\beta\in\{y,z\}$ and measurement along direction $\alpha\in\{y,z\}$ are shown in Fig.~\ref{fig:dephasing-correlators} for typical experimental parameters. Here we account for both pure dephasing from electric-field fluctuations and modulations of the decay envelope due to hyperfine coupling. Notably, $C_{yy}$ and $C_{zz}$ show a strong full-amplitude decay with small modulations [Figs.~\ref{fig:dephasing-correlators}(a) and \ref{fig:dephasing-correlators}(b)]. In contrast, $C_{zy}$ and $C_{yz}$ [Figs.~\ref{fig:dephasing-correlators}(c) and \ref{fig:dephasing-correlators}(d)] \emph{grow} on a very short time scale on the order of the inverse hole-spin precession frequency, and subsequently slowly decay. Within the approximations made above, there will generally be a small non-decaying portion of the coherence arising from counter-rotating contributions to $C_{\varphi}$ that are independent of the fluctuating Zeeman energy to leading order. 

\begin{figure}
\includegraphics[width = 0.45\textwidth]{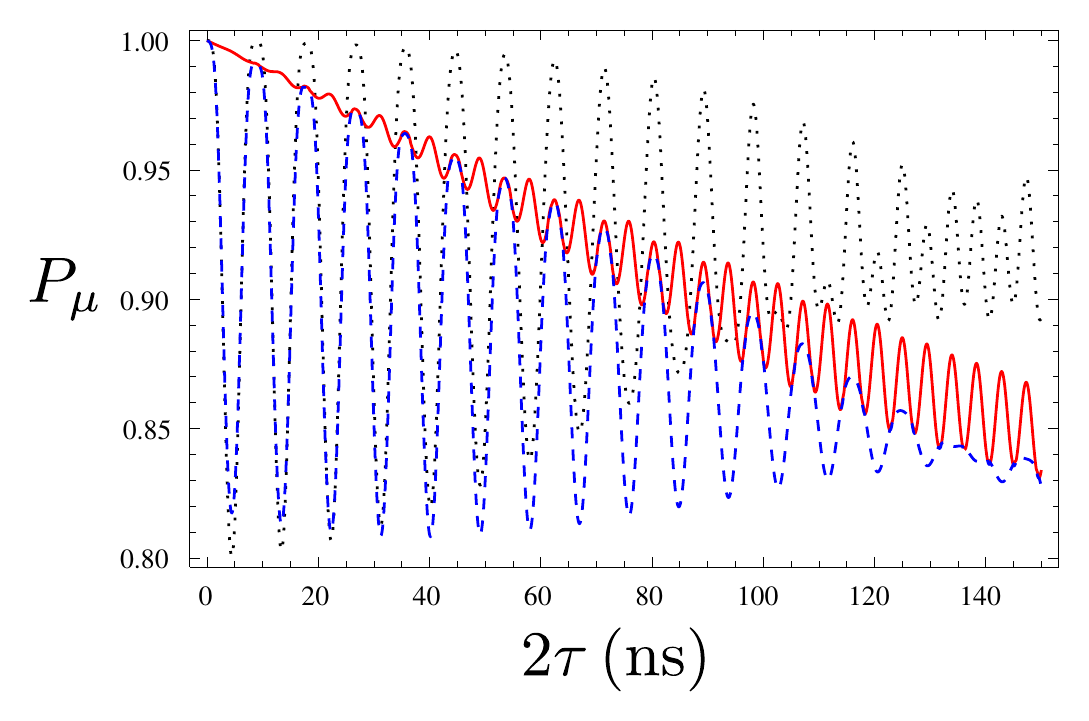}
\caption{(Color online) Purities for preparation along the unit vectors $\mathbf{\hat{e}}_\mu$ in the presence of a Markovian pure dephasing process giving $T_\phi=1\,\mu\mathrm{s}$. The purities are $P_-$ (red solid line), $P_+$ (blue dashed line), and $P_x$ (black dotted line). The material parameters are as in Figs.~\ref{fig:correlators-rotating} and \ref{fig:spinpurity} for an $\mathrm{In_{0.5}Ga_{0.5}As}$ quantum dot, and for this plot we have taken a magnetic field of $B=400\,\mathrm{mT}$. In the presence of the pure-dephasing process, at certain times it becomes advantageous to prepare the qubit along the $\mathbf{\hat{x}}$ direction [when $P_x(2\tau)>P_-(2\tau)$].}
\label{fig:dephasing-purities-all}
\end{figure}

Detail of the purities for initialization along each of the three principal directions $(\mathbf{\hat{e}}_\pm,\mathbf{\hat{e}}_x)$ is shown in Fig.~\ref{fig:dephasing-purities-all}, accounting for pure dephasing. For time $\tau\lesssim \tau_c$, with $\tau_c$ given by Eq.~\eqref{eq:PureDephasingTime}, the optimal initialization axis alternates between $\mathbf{\hat{e}}_-$ (giving $P_-$, i.e., initialization perpendicular to the magnetic-field quantization axis) and $\mathbf{\hat{x}}$ (giving $P_x$, initialization along the magnetic field). In contrast, for $\tau >\tau_c$, pure-dephasing processes dominate over the effect of envelope modulations and it will be advantageous once again to initialize a hole spin along the magnetic field.

\begin{figure}
\includegraphics[width = 0.45\textwidth]{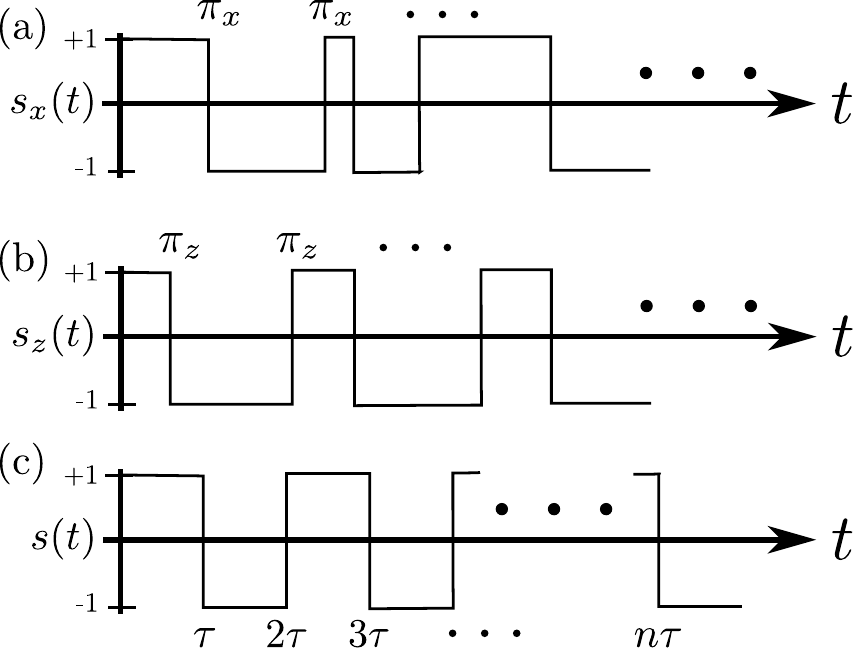}
\caption{Sign functions for (a) $\pi_x$-pulses, (b) $\pi_z$-pulses, and (c) for a periodic dynamical decoupling (PDD) sequence consisting of $n$ equally-spaced $\pi$-pulses (shown here for $n$ odd).}
\label{fig:PulsesSchematic}
\end{figure}

\section{Two-axis dynamical decoupling}\label{sec:DynamicalDecoupling}

As is well-known, a sequence of many $\pi$ rotations applied in rapid succession can be used to decouple a qubit from an environment having a finite correlation time, by averaging the interaction to zero.\cite{Viola1998,Khodjasteh2005,uhrig2007} In general, to simultaneously control fluctuations along the magnetic-field axis (due, e.g., to $g$-factor modulation) and transverse to the magnetic-field axis (due, e.g., to hyperfine coupling), it is useful to consider $\pi$ rotations about two orthogonal axes.\cite{Viola1999} Rotations about the $\hat{\mathbf{x}}$-axis ($\pi_x$-pulses) lead to $S_z\to -S_z$, averaging out the Ising-like hyperfine coupling $\sim h_zS_z$. Rotations about the $\hat{\mathbf{z}}$-axis ($\pi_z$-pulses) result in $S_x\to -S_x$, averaging out the Zeeman term, $\sim \gamma_H B S_x$. We can generally account for a sequence of fast $\pi_x$- and $\pi_z$-pulses with the replacements:
\begin{eqnarray}
H_0 &\to & H_0(t) = H_S(t)+H_E,\\
V(t) &\to & V(t) = s_x(t)H_\mathrm{hf},
\end{eqnarray} 
where $s_x(t)$ is the sign function for $\pi_x$-pulses. The system Hamiltonian $H_S(t)$ generally accounts for a time-dependent fluctuating Zeeman splitting and a sign function for $\pi_z$-pulses, $s_z(t)$: 
\begin{equation}
H_S(t) = -s_z(t)\left[\gamma_H B+\delta\omega(t) \right]S_x.
\end{equation}
\begin{figure}
\includegraphics[width = 0.45\textwidth]{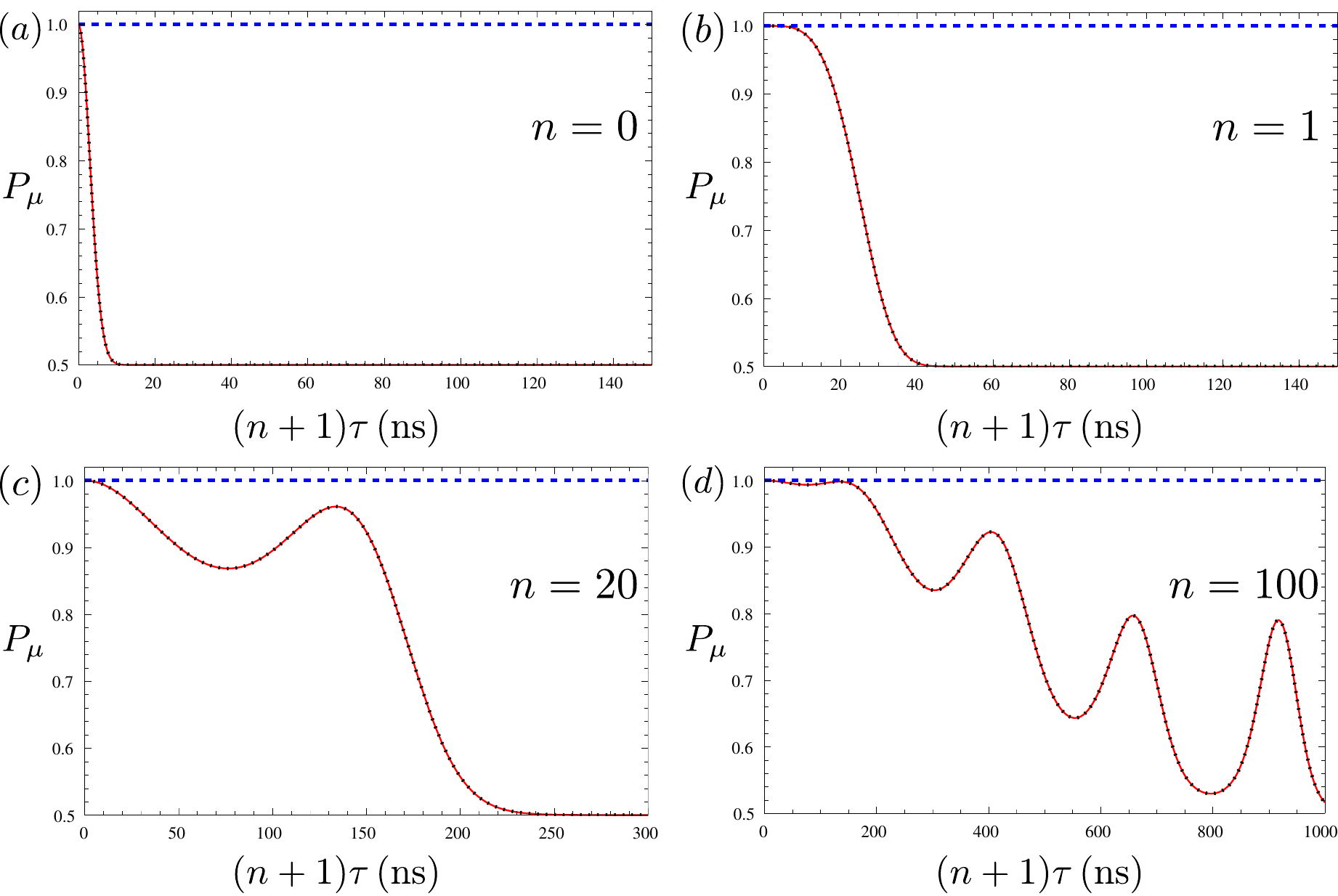}
\caption{(Color online) Evolution of hole-spin purity under an $n$-pulse periodic dynamical decoupling (PDD) sequence $\left[\left(\tau-\pi\right)^n-\tau\right]$ with $g_\perp=\gamma_H=0$. In this case, the purities are identical for $\pi$-rotations about $\hat{\mathbf{x}}$ ($\pi=\pi_x$) or $\hat{\mathbf{y}}$ ($\pi=\pi_y$). The purities are $P_-$ (blue dashed line), $P_+$ (red solid line), and $P_x$ (black dotted line), for initialization along $\mathbf{\hat{e}}_-$, $\mathbf{\hat{e}}_+$, and $\mathbf{\hat{e}}_x$, respectively. We show dynamics for (a) free-induction decay, $n=0$, (b) Hahn echo, $n=1$, (c) $n=20$, and (d) $n=100$. We have assumed the same material parameters as in Figs.~\ref{fig:correlators-rotating} and \ref{fig:spinpurity} for an $\mathrm{In_{0.5}Ga_{0.5}As}$ quantum dot, but here we assume a magnetic field of $B=400\,\mathrm{mT}$. Recurrences occur with a period $\sim 2\pi\hbar/\gamma_\mathrm{In}B\simeq 266\,\mathrm{ns}$ (given by the indium Larmor frequency), with the first maximum at $\sim \pi\hbar/\gamma_\mathrm{In}B$ for $n$ even.}
\label{fig:PDD-no-gammaH}
\end{figure}
This leads directly to the complex-valued filter functions
\begin{equation}\label{eq:TwoAxisFilter}
Z_{j\pm}(t) = \sigma_j\int_0^t dt' s_x(t')e^{i\left[\phi_z(t')\pm\gamma_j Bt'\right]},
\end{equation}
with
\begin{equation}
\phi_z(t) = \int_0^t dt' s_z(t')\left[\gamma_H B+\delta\omega(t')\right].
\end{equation}
Equation \eqref{eq:TwoAxisFilter} can now be substituted into the previous expressions to find the purity and associated principal axes for an arbitrary interlaced sequence of $\pi_x$- and $\pi_z$-pulses [see Fig.~\ref{fig:PulsesSchematic}(a,b)]. 

\begin{figure}
\includegraphics[width = 0.45\textwidth]{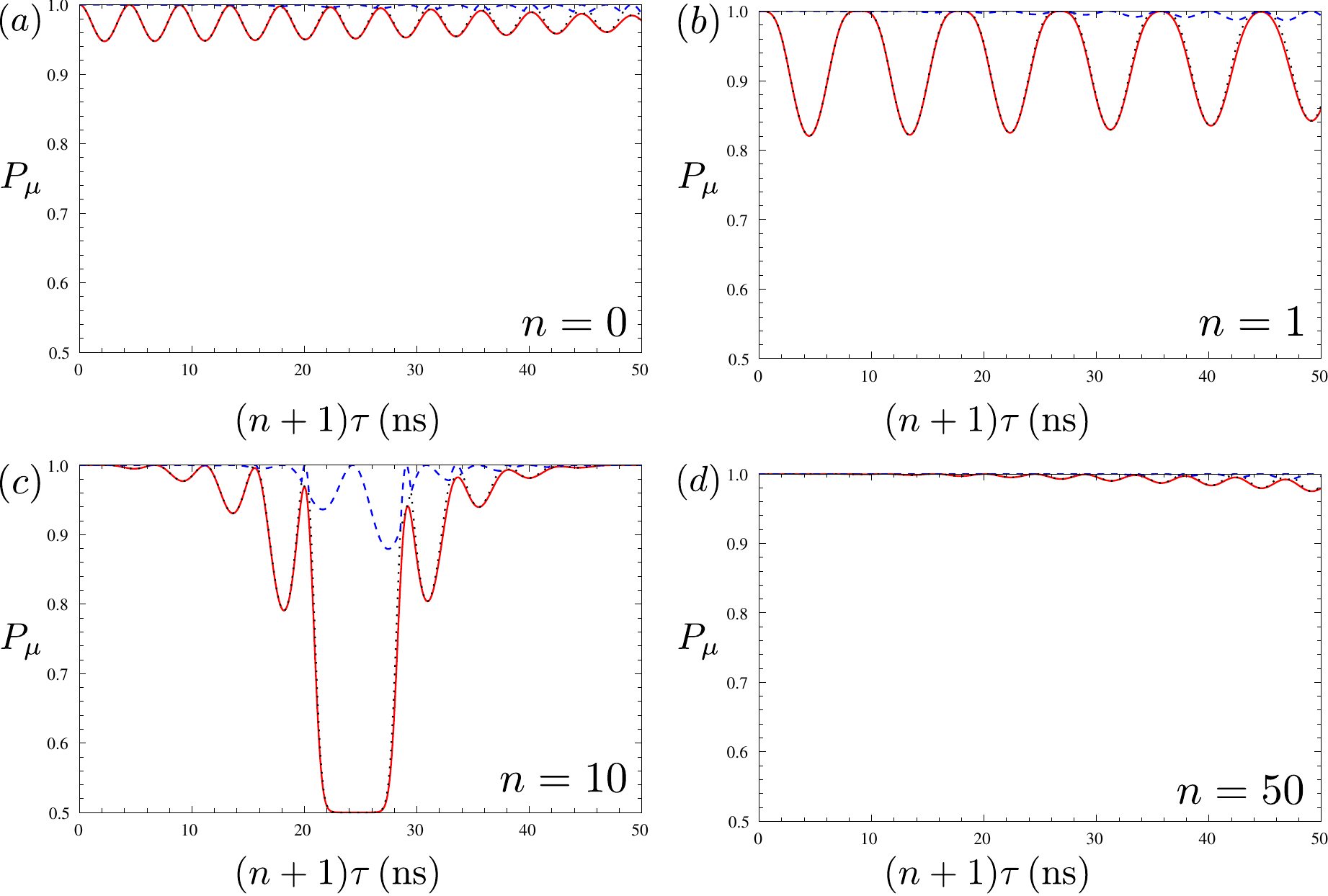}
\caption{(Color online) Hole-spin purity under an $n$-pulse periodic dynamical decoupling (PDD-X) sequence $\left[\left(\tau-\pi_x\right)^n-\tau\right]$. Here, we take $g_\perp=0.04$, but all other parameters are equivalent to those given in the caption of Fig.~\ref{fig:PDD-no-gammaH}. The purities are $P_-$ (blue dashed line), $P_+$ (red solid line), and $P_x$ (black dotted line), for initialization along $\mathbf{\hat{e}}_-$, $\mathbf{\hat{e}}_+$, and $\mathbf{\hat{e}}_x$, respectively. We show dynamics for (a) free-induction decay, $n=0$, (b) Hahn echo, $n=1$, (c) $n=10$, and (d) $n=50$. Resonances occur [as in (c)] when the frequency of $\pi_x$-pulses is comparable to the hole-spin precession frequency.}
\label{fig:PDDX}
\end{figure}

\begin{figure}
\includegraphics[width = 0.45\textwidth]{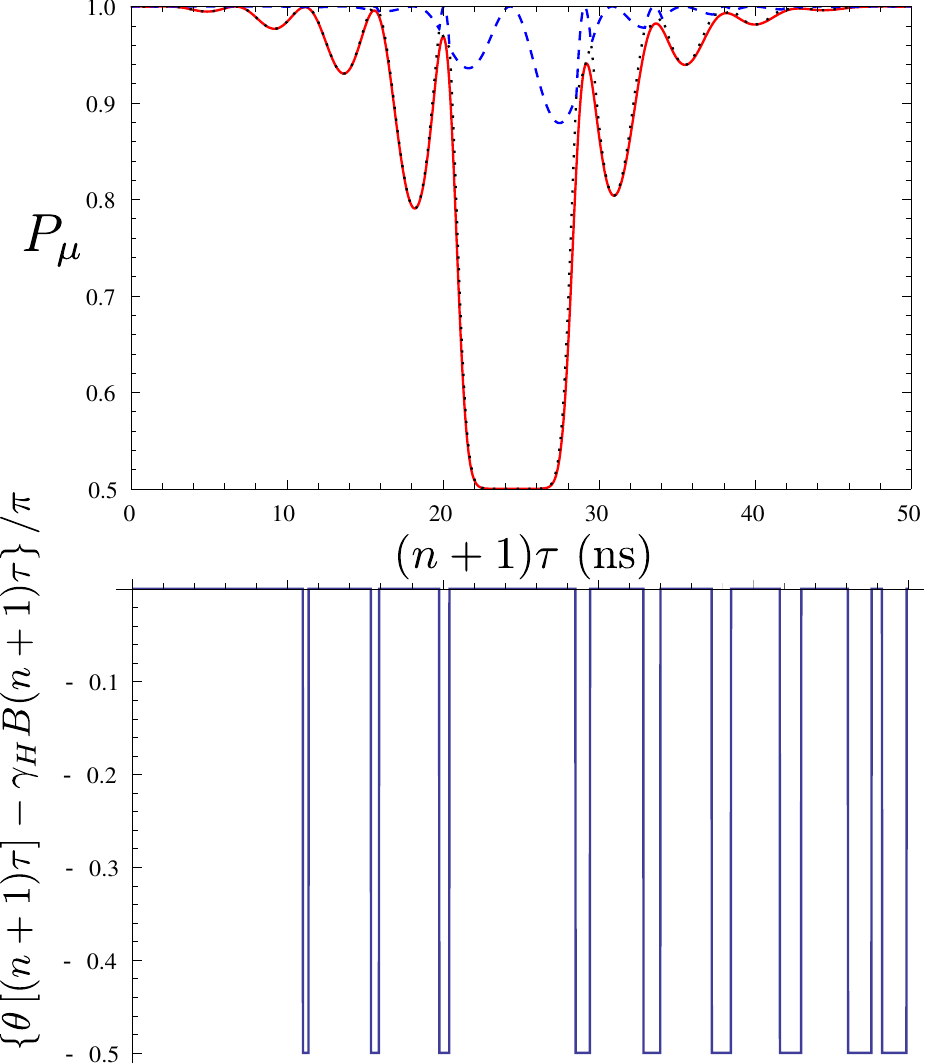}
\caption{(Color online) Purities (top) and angle $\theta[(n+1)\tau]$ (bottom) defining principal axes for a PDD-X sequence with $n=10$, using the same parameters and line labeling as Fig.~\ref{fig:PDDX}(c).}
\label{fig:PDDX-phases}
\end{figure}

\begin{figure}
\includegraphics[width = 0.45\textwidth]{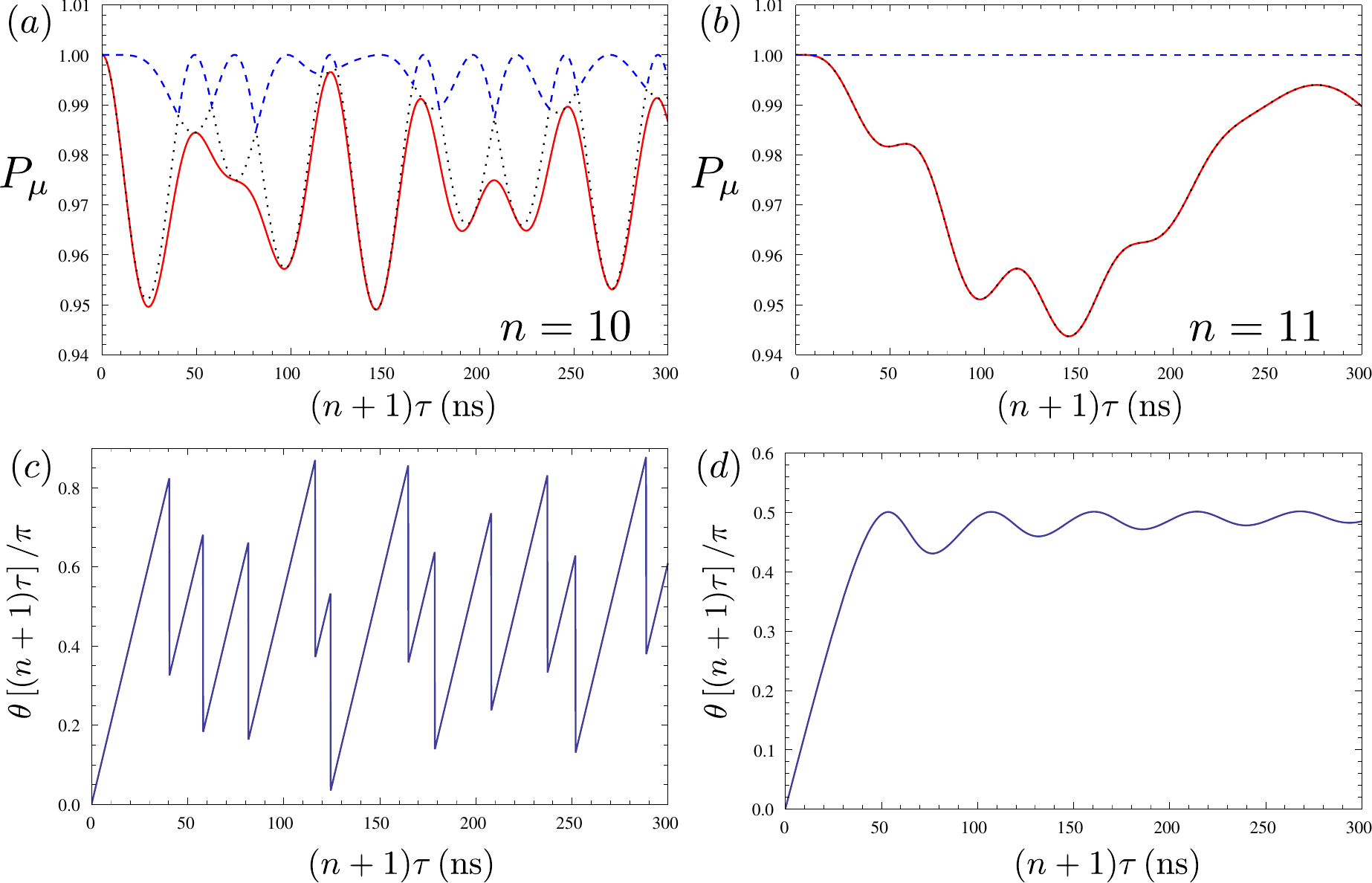}
\caption{(Color online) Purities $P_\mu$ [(a), (b)] and associated phase $\theta$ [(c), (d)] defining principal axes for an $n$-pulse PDD-Y sequence $\left[\left(\tau-\pi_y\right)^n-\tau\right]$, with $n=10$ (even) [(a) and (c)] and $n=11$ (odd) [(b) and (d)]. Line styles and parameters are the same as in Figs.~\ref{fig:PDDX} and \ref{fig:PDDX-phases}}
\label{fig:PDDY}
\end{figure}

In this section, for simplicity, we will assume negligible noise in the Zeeman splitting [$\delta\omega (t)\simeq 0$]. Further, we will focus on two specific (simple) dynamical decoupling sequences: Periodic dynamical decoupling with equally spaced $\pi_x$-pulses (PDD-X) and equally spaced $\pi_y$-pulses (PDD-Y), for which
\begin{eqnarray}
s_x(t) = s(t),\quad s_z(t)=1\quad (\mathrm{PDD}\mbox{-}\mathrm{X}),\\
s_x(t)=s_z(t)=s(t)\quad (\mathrm{PDD}\mbox{-}\mathrm{Y}),
\end{eqnarray}
where
\begin{equation}\label{eq:PDDSignFunction}
s(t) = 1+2\sum_{k=1}^n(-)^k\theta(t-k\tau).
\end{equation}
Equation \eqref{eq:PDDSignFunction} is illustrated schematically in Fig.~\ref{fig:PulsesSchematic}(c). In this case, it is straightforward to evaluate Eq.~\eqref{eq:TwoAxisFilter} analytically. We give explicit analytical forms for $Z_{j\pm}(t)$ in Appendix \ref{sec:DDFilterFunctions}. The resulting purity decay and associated angle $\theta$ determining the principal axes are shown for a range of parameters in Figs.~\ref{fig:PDD-no-gammaH}-\ref{fig:PDDY}.

Figure \ref{fig:PDD-no-gammaH} illustrates purity decay for the case of a vanishing hole-spin $g$-factor, $\gamma_H B\to 0$. In this limit, the filter functions are given [see Eq.~\eqref{eq:PDDXFilter}] by
\begin{equation}
Z_{j\pm} = \frac{\sigma_j}{\gamma_j B}\tan\left(\frac{\gamma_jB\tau}{2}\right)\left[1+(-)^n e^{\pm i\gamma_j B(n+1)\tau}\right].
\end{equation}
In this case, all fluctuations $\tilde{h}^z(t)$ are along $\hat{\mathbf{z}}$. Due to rotational symmetry about $\hat{\mathbf{z}}$, the magnitude of $S_z$ is preserved for all time (blue dashed line in Fig.~\ref{fig:PDD-no-gammaH}), and the dynamics are generally identical for repeated $\pi_x$-pulses (PDD-X) or repeated $\pi_y$-pulses (PDD-Y). A spin prepared along any other axis will decay with partial recurrences near the zeroes of $|Z_{j\pm}|$. These are separated by the typical nuclear-spin precession period, $\sim 2\pi/\gamma_j B$. The first zero after $\tau\ne 0$ occurs at $(n+1)\tau\simeq \pi/\gamma_j B$ for $n$ even and at $(n+1)\tau\simeq 2\pi/\gamma_j B$ for $n$ odd.

With a nonzero hole-spin $g$-factor, the purity dynamics depend strongly on the decoupling sequence (PDD-X or PDD-Y). Dynamics for a PDD-X sequence are shown in Fig.~\ref{fig:PDDX} with typical parameters for a heavy-hole spin in a quantum dot. At a critical time scale, a resonant dip develops in the purity dynamics [see Fig.~\ref{fig:PDDX}(c)]. This dip is a consequence of the well-known phenomenon of accelerated decoherence\cite{Viola1998} and can be understood from the filter functions reported in Appendix \ref{sec:DDFilterFunctions}, giving: 
\begin{equation}
Z_{j\pm}[(n+1)\tau]\simeq i \frac{2\sigma_j (n+1)}{\omega_{j\pm}},\quad\omega_{j\pm}\tau\to\pi. 
\end{equation} 
Thus, the degree of purity decay $\sim |Z_{j\pm}|^2$ is bounded but increasing for small $n$. The absolute time scale for the dip, $(n+1)\tau\simeq (n+1)\pi/\omega_{j\pm}$, can be pushed out to longer time by increasing $n$. This resonant dip is similar to that identified as a useful tool for sensing.\cite{Zhao2011,Zhao2012} The methods presented here can be used to preserve pure qubit states in spite of these resonant dips [blue dashed curve in Fig.~\ref{fig:PDDX}(c)], when it is not possible to suppress these dips with faster $\pi$-pulses [Fig.~\ref{fig:PDDX}(d)]. Alternatively, this method can be used to identify the initialization direction that would be most susceptible to purity decay, enhancing signal-to-noise when such a resonant dip is used for sensing. In Fig.~\ref{fig:PDDX-phases}, we show the evolution of the angle $\theta$ defining principal axes near the resonant dip shown in Fig.~\ref{fig:PDDX}(c). 

Resonant dips such as those shown in Fig.~\ref{fig:PDDX}(c) can be avoided altogether within this model by performing a sequence of repeated $\pi$-pulses about the $y$-axis (PDD-Y). Evolution under an $n$-pulse PDD-Y sequence is shown for a heavy-hole spin in a quantum dot in Fig. \ref{fig:PDDY}. For $n$ even [Figs.~\ref{fig:PDDY}(a,c)], phase evolution is not symmetric about the halfway point, $(n+1)\tau/2$, leading to nontrivial jumps in the purity evolution and associated angle $\theta$. In contrast, $n$ odd [Fig.~\ref{fig:PDDY}(b,d)] allows for symmetric time-reversed dynamics, unwinding phase evolution under the Zeeman term. This distinction between time-symmetric and time-asymmetric decoupling sequences is well known.\cite{Faoro2004} For a PDD-Y sequence with $n$ odd, we find $|Z_{j+}|=|Z_{j-}|$ [see Appendix \ref{sec:DDFilterFunctions}], leading to $\lambda_-=0$ [see Eq.~\eqref{eq:lambdapm}]. Thus, to leading order in the Magnus expansion, the purity can be preserved perfectly with the correct initialization [blue dashed line in Fig.~\ref{fig:PDDY}(b)].

In the limit of an $S$-$T_0$ qubit ($\gamma_j\to 0$), a PDD-Y sequence with $n$ odd leads to $Z_{j\pm}=0$ [from Eq.~\eqref{eq:PDDYFilter}], giving no decay for any initialization direction. As pointed out in Ref.~\onlinecite{Bergli2007} for the analogous problem of a Josephson charge qubit coupled to two-level fluctuators, this result actually holds to all orders in a Magnus expansion. Of course, pure dephasing due to exchange fluctuations would lead to a finite decay even in this case.

\section{Conclusions}\label{sec:Conclusion}

We have given a general procedure for the calculation of the non-Markovian dynamics of qubit purity for qubits interacting with an anisotropic environment. Applying this procedure to the case of a hole-spin or $S$-$T_0$ qubit interacting with a nuclear-spin bath, we find that (at sufficiently short times), the qubit purity is maximized by storing the qubit in a superposition of non-interacting eigenstates. Storage of the qubit in its non-interacting ground state can actually be the \emph{worst} choice for these systems. That storage in the computational basis (non-interacting eigenbasis) is sub-optimal is not unique to hole spins and $S$-$T_0$ qubits. We expect this to be true for a wide variety of qubit systems when ancillas are required a short time after preparation and if pure-dephasing processes are weak. This effect is especially pronounced for systems interacting with anisotropic non-Markovian environments, including hole spins, nitrogen-vacancy center spins, and spins bound to phosphorus donor impurities.

In the process of calculating purity for a hole-spin qubit, we have given closed-form analytical expressions for all spin components describing the spin-echo and dynamical-decoupling dynamics of hole spins in the presence of a nuclear-spin bath. In particular, we have shown how echo envelope modulations can be described by a combination of terms arising from (i) a rotation to a set of principal axes $\mathbf{\hat{e}}_\mu$ for the generator of evolution $\mathcal{L}$, and (ii) modulations in a set of eigenvalues $\lambda_\mu$. While both contributions enter into the spin dynamics in general, the eigenvalues are most important for determining the purity, provided the spin is initialized along an appropriate principal axis. We have fully accounted for a pure-dephasing process arising from white-noise fluctuations in the hole-spin Zeeman energy and have illustrated the resulting rich dynamics. All of the results presented here are directly applicable to $S$-$T_0$ qubits, under the mapping described in Sec.~\ref{sec:SingletTripletMapping}.

We expect the calculations for qubit purity given here to be useful in quantum-information protocols that require high-purity ancillas, including quantum error correction, algorithmic cooling, and methods for high-fidelity readout. The general approach taken here emphasizes the fact that, for anisotropic systems, optimizing coherence is not simply a matter of manipulating the \emph{spectral content} of the noise [associated with eigenvalues $\lambda_\mu$], but also the \emph{geometry} of the noise, determined by initializing with respect to principal axes $\mathbf{\hat{e}}_\mu$.

\begin{acknowledgments}
We thank L. Childress for very useful discussions. WAC and SC thank the KITP China for their generous hospitality where some of this work was completed. We acknowledge financial support from NSERC, CIFAR, INTRIQ, and FRQNT. SC acknowledges support from the Chinese Youth 1000 Talents Program.
\end{acknowledgments}

\appendix

\section{Average Hamiltonian and generator} \label{sec:approx_gneq0}

We take the leading-order Magnus Hamiltonian to have the general form
\begin{equation}
H^{(0)}(t) = \int_0^t dt'\tilde{V}(t') = \sum_\alpha \mathcal{B}_\alpha (t)S_\alpha,\\
\end{equation}
where $\mathcal{B}_\alpha(t)$ are Hermitian bath operators that act exclusively on the environment.

The matrix elements of $\mathcal{L}_0$ [defined by Eq.~\eqref{eq:LApprox} of the main text] can then be written in terms of $\boldsymbol{\mathcal{B}}=\left(\mathcal{B}_x,\mathcal{B}_y,\mathcal{B}_z\right)^T$ as
\begin{eqnarray}
\left[\mathcal{L}_0\right]_{0\alpha} & = & \frac{1}{2}\mathrm{Tr}\left\{\sigma_0\mathcal{L}_0 S_\alpha\right\}\\
&=&-\frac{1}{2}\mathrm{Im}\left<\boldsymbol{\mathcal{B}}\times\boldsymbol{\mathcal{B}}\right>_E\cdot\mathbf{\hat{e}_\alpha},\label{eq:L0Ialpha}\\ 
\left[\mathcal{L}_0\right]_{\alpha\beta} & = & 2\mathrm{Tr}\left\{S_\alpha \mathcal{L}_0 S_\beta\right\}\\
&=& \mathrm{Re}\left\{\delta_{\alpha\beta}\left<\boldsymbol{\mathcal{B}}\cdot\boldsymbol{\mathcal{B}}\right>_E-
\left<\mathcal{B}_\beta \mathcal{B}_\alpha\right>_E\right\}.\label{eq:L0alphabeta}
\end{eqnarray}
Here, $S_\alpha = \sigma_\alpha/2$ are spin-1/2 operators for $\alpha=x,y,z$, while $\sigma_0$ is the identity in the qubit Hilbert space and $\mathbf{\hat{e}}_\alpha$ is a unit vector along an axis in Cartesian coordinates, ($\mathbf{\hat{e}}_x=\mathbf{\hat{x}}$, $\mathbf{\hat{e}}_y=\mathbf{\hat{y}}$, $\mathbf{\hat{e}}_z=\mathbf{\hat{z}}$). In Eqs.~\eqref{eq:L0Ialpha} and \eqref{eq:L0alphabeta}, we have used the fact that the bath operators are Hermitian, giving $\mathrm{Im}\left<\mathcal{B}_\alpha\mathcal{B}_\beta\right>=\left<\left[\mathcal{B}_\alpha,\mathcal{B}_\beta\right]\right>/2$ and $\mathrm{Re}\left<\mathcal{B}_\alpha\mathcal{B}_\beta\right>=\left<\left\{\mathcal{B}_\alpha,\mathcal{B}_\beta\right\}\right>/2$, where $[,]$ indicates a commutator and $\{,\}$ is an anticommutator. A sufficient condition for the inhomogeneous term, Eq.~\eqref{eq:L0Ialpha}, to vanish within a leading-order Magnus expansion, Eq.~\eqref{eq:InhomogeneousZero}, is then:
\begin{equation}
\mathrm{Im}\left<\boldsymbol{\mathcal{B}}\times\boldsymbol{\mathcal{B}}\right>_E=0\Rightarrow \overline{\left<\delta\mathbf{S}(t)\right>}=0.
\end{equation}

We now consider the most general anisotropic hyperfine Hamiltonian,
\begin{equation}
V = H_\mathrm{ahf} = \sum_{\gamma\delta,k}A^{\gamma\delta}_k I_k^\gamma S_\delta.
\end{equation} 
If the environment Hamiltonian is described by a general inhomogeneous Zeeman term,
\begin{equation}
H_I = -\sum_k\gamma_k \mathbf{B}_k\cdot\mathbf{I}_k,
\end{equation}
the interaction picture results in a rotation
\begin{equation}\label{eq:Bdelta}
\mathcal{B}_\beta(t) = \sum_{k,\gamma\delta\alpha} A_k^{\gamma\delta}g_k^{\gamma\delta,\alpha\beta}(t) I_{k}^\alpha.
\end{equation}
For the case of nuclear spin $I=1/2$, for example, the coefficients in the expansion of Eq.~\eqref{eq:Bdelta} are given explicitly by
\begin{equation}
g_{k}^{\gamma\delta,\alpha\beta}(t) = \int_0^t dt'2\mathrm{Tr}\left\{ \tilde{I}_k^\gamma(t')I_{k}^\alpha\right\}\cdot 2\mathrm{Tr}\left\{ \tilde{S}_\delta(t')S_\beta\right\}.
\end{equation}
With Eq.~\eqref{eq:Bdelta}, it is straightforward to estimate the matrix elements given in Eq.~\eqref{eq:L0Ialpha}. All terms are proportional to the initial polarization of the nuclear-spin system and therefore vanish:
\begin{equation}
\mathrm{Im}\left<\boldsymbol{\mathcal{B}}\times\boldsymbol{\mathcal{B}}\right>_I =0. 
\end{equation}
Here, we have used the subscript $I=E$ for the nuclear-spin environment. Thus, for an initially unpolarized nuclear spin bath, we are justified in neglecting the inhomogeneous term to leading order in a Magnus expansion,
\begin{equation}
\left<I^\gamma_k\right>_I=0\Rightarrow \overline{\left<\delta\mathbf{S}(t)\right>}\simeq 0.
\end{equation} 
Specializing to the case of an Ising-like hyperfine interaction,
\begin{equation}
A_k^{\alpha\beta} = \delta_{\alpha z}\delta_{\beta z} A_k,
\end{equation} 
and the spin-echo problem discussed in Sec.~\ref{sec:model}, we find explicit forms for the bath operators $\mathcal{B}_\alpha$, in terms of the complex-valued filter functions $Z_{j\pm}(2\tau)$ given in Eq.~\eqref{eq:ZiDefinition} of the main text:
\begin{eqnarray}
\mathcal{B}_x & = & 0,\\
\mathcal{B}_y & = & \sum_j\frac{1}{4\sigma_j}\left[\left(Z_{j-}-
Z_{j+}^*\right)h_j^+-\left(Z_{j+}-Z_{j-}^*\right)h_j^-\right],\nonumber\\ 
\mathcal{B}_z & = & \sum_j\frac{-i}{4\sigma_j}\left[\left(Z_{j-}+
Z_{j+}^*\right)h_j^+-\left(Z_{j+}+Z_{j-}^*\right)h_j^-\right].\nonumber 
\end{eqnarray}
Applying the rules in Eq.~\eqref{eq:sigma} for an uncorrelated and unpolarized nuclear-spin state immediately gives the non-vanishing correlators,
\begin{eqnarray}
\left<\mathcal{B}_y\mathcal{B}_y\right>&=&\frac{1}{2}\left(\lambda_x-\mathrm{Re}Z^2\right),\\
\left<\mathcal{B}_z\mathcal{B}_z\right>&=&\frac{1}{2}\left(\lambda_x+\mathrm{Re}Z^2\right),\\
\left<\mathcal{B}_y\mathcal{B}_z\right>&=&\left<\mathcal{B}_z\mathcal{B}_y\right>=-\frac{1}{2}\mathrm{Im}Z^2.
\end{eqnarray}
Here, $Z^2(2\tau)$ and $\lambda_x(2\tau)$ are given in Eqs.~\eqref{eq:ZDefinition} and \eqref{eq:lambdax} of the main text, respectively. Inserting these correlators $\left<\mathcal{B}_\alpha \mathcal{B}_\beta\right>$ into Eq.~\eqref{eq:L0alphabeta} directly gives the matrix form found in the main text [Eqs.~\eqref{eq:liouvillian} and \eqref{eq:LyzDefinition}].

Parenthetically, we note that \emph{arbitrary} initial conditions for the bath can be taken above, in principle, including pure-state environment initial conditions (required to measure entanglement through purity). However, for each bath initial state, it will be important to justify the Gaussian approximation used to derive Eq.~\eqref{eq:LApprox}. This approximation is very good for an uncorrelated thermal bath or a sufficiently random `narrowed' state,\cite{Beaudoin2013} but may break down for pure initial conditions with strong (classical or quantum) correlations. 

\section{Simple example: $\gamma_j=0$} \label{sec:SimpleExample}

It is useful to consider a simple and direct application of the analytical expressions derived in Appendix \ref{sec:approx_gneq0}. Here we consider dynamics at a time scale short compared to the nuclear-spin precession period, and neglect the nuclear gyromagnetic ratio, so there is effectively one nuclear-spin species $j$ with $\gamma_j\simeq 0$ and we assume a single nuclear-field variance $\sigma_j=\sigma_N$. This limit is directly applicable to $S$-$T_0$ qubits (see Sec.~\ref{sec:SingletTripletMapping}) Under these conditions, there is only one complex filter function, $Z_{j\pm}(2\tau) = Z(2\tau)$, for one fixed $j$ [see Eq.~\eqref{eq:ZiDefinition}]. Setting $\omega=\gamma_H B$,
\begin{equation}
Z(2\tau) = \frac{4\sigma_N}{\omega}\sin^2\frac{\omega\tau}{2} e^{i\left(\omega\tau-\pi/2\right)}.
\end{equation} 

From Eq.~\eqref{eq:LyzDefinition}, we see that the submatrix can be rewritten as an outer product (a projector onto a vector constructed from the real and imaginary parts of $Z=X+iY$):
\begin{equation}
\left[\mathcal{L}^{yz}_0(2\tau)\right] = \begin{pmatrix}X^2 & XY\\ XY & Y^2\end{pmatrix}=\begin{pmatrix}X\\ Y\end{pmatrix}\begin{pmatrix}X & Y\end{pmatrix}.
\end{equation}
The eigenvectors giving $\mathbf{\hat{e}}_\pm$ are then simply this vector and the vector orthogonal to it:
\begin{eqnarray}
\left[\mathcal{L}^{yz}_0(2\tau)\right]\begin{pmatrix}X\\Y\end{pmatrix}&=&|Z|^2\cdot \begin{pmatrix}X\\Y\end{pmatrix},\label{eq:SubmatrixEplus}\\
\left[\mathcal{L}^{yz}_0(2\tau)\right]\begin{pmatrix}Y\\-X\end{pmatrix}&=&0.\label{eq:SubmatrixEminus}
\end{eqnarray} 
In terms of unit vectors, 
\begin{eqnarray}
\mathbf{\hat{e}}_+(2\tau) & = & \frac{1}{|Z(2\tau)|}\left(X(2\tau)\mathbf{\hat{y}}+Y(2\tau)\mathbf{\hat{z}}\right),\label{eq:eplusZ}\\
\mathbf{\hat{e}}_-(2\tau) & = & \frac{1}{|Z(2\tau)|}\left(-Y(2\tau)\mathbf{\hat{y}}+X(2\tau)\mathbf{\hat{z}}\right).\label{eq:eminusZ}
\end{eqnarray}
Equation \eqref{eq:eplusZ} gives a quick shortcut to find the angle $\theta(2\tau)$ in Fig.~\ref{fig:eigens}:
\begin{equation}\label{eq:ThetaExample}
\theta(2\tau) = \arg Z(2\tau) = \omega\tau -\frac{\pi}{2}.
\end{equation}
From Eqs.~\eqref{eq:SubmatrixEplus} and \eqref{eq:SubmatrixEminus}, we can read off the eigenvalues,
\begin{eqnarray}
\lambda_+(2\tau) &=& |Z(2\tau)|^2=\left(\frac{4\sigma_N}{\omega}\right)^2 \sin^4\frac{\omega\tau}{2}, \label{eq:lambda-plus-modulation}\\
\lambda_-(2\tau) &=& 0.\label{eq:lambda-minus-zero}
\end{eqnarray}

Equation \eqref{eq:lambda-minus-zero} indicates that a spin initially aligned along $\mathbf{\hat{e}}_-$ will show no decay or modulations under a spin-echo sequence within the range of applicability of approximations made here. On the surface, this may not seem surprising since we have assumed $\gamma_j=0$, making the bath static and the spin-echo dynamics reversible. However, a spin prepared along any other direction will show violent modulations, as described by Eq.~\eqref{eq:lambda-plus-modulation}. The distinction between these two cases can be understood by considering the specific geometry and the phase in Eq.~\eqref{eq:ThetaExample}. Before the first $\pi$-pulse, a spin initialized along $\mathbf{\hat{e}}_+$ will evolve with a phase $\phi(t)$ [see also Eq.~\eqref{eq:SPrime} for an analogous expression after the echo sequence has been carried out]:
\begin{equation}
\left<S_+^\prime(t)\right>\propto e^{i\phi(t)},\quad \phi(t) = -\omega t+\theta(2\tau).
\end{equation}
Note that $\theta(2\tau)$ does not evolve with $t$ since it determines the initial condition. Inserting Eq.~\eqref{eq:ThetaExample}, we see that the initial condition is such that the spin lies along $-\mathbf{\hat{z}}$ at the time of the first $\pi$-pulse ($t=\tau$):
\begin{equation}
\phi(\tau) = -\frac{\pi}{2}.
\end{equation}
This situation leads to rapid envelope modulations. In contrast, a spin initialized along $\mathbf{\hat{e}}_-$ will be oriented along $\mathbf{\hat{y}}$ at the first $\pi$-pulse and will show no modulations. We can understand this difference by considering a model of a spin evolving in the presence of a classical magnetic field, $\mathbf{B} = B_x\mathbf{\hat{x}}+\delta B_z{\mathbf{\hat{z}}}$, having a fixed $x$-component $B_x$ and slowly-varying random $z$-component $\delta B_z$. For initialization along $\mathbf{\hat{e}}_-$, a finite $\delta B_z$ will result in a finite component along $\mathbf{\hat{x}}$ at the time of the first $\pi$-pulse, but the spin will lie approximately in the $y$-$x$ plane due to the choice of initial condition (with small corrections in $\delta B_z/B_x\ll 1$). In this plane, the system shows perfect mirror symmetry for a reflection through the $x$-axis, so a $\pi$-pulse about $\mathbf{\hat{x}}$ induces symmetric time-reversed dynamics, returning the spin precisely to its starting point in the rotating frame after a second $\pi$-pulse is performed at $t=2\tau$. In contrast, if the spin is initialized along $\mathbf{\hat{e}}_+$, it will lie approximately in the $x$-$z$ plane at the time of the first $\pi$-pulse. In this plane, for any finite value of $\delta B_z$, there is no reflection symmetry for a $\pi$-rotation about the $x$-axis. The spin's cone of precession after the $\pi$-pulse can be quite different from that before the $\pi$-pulse, resulting in a mismatch in evolutions causing the modulations indicated by Eq.~\eqref{eq:lambda-plus-modulation} for any finite $\delta B_z$. 

\section{Averages for pure dephasing}\label{sec:AvgPureDephasing}

Here we give expressions for the averages required to evaluate the associated generator $\left<\mathcal{L}_0\right>_\phi$, accounting for averages over realizations of the Gaussian random variable $\delta\omega(t)$ described by Eq.~\eqref{eq:delta-omega}.

Explicitly, the eigenvalue $\left<\lambda_x\right>_\phi$ can be written as
\begin{equation}
\langle\lambda_x\rangle_\phi = \frac{1}{2}\sum_j\left(\langle|Z_{j+}|^2\rangle_\phi+\langle|Z_{j-}|^2\rangle_\phi\right),
\end{equation}
where
\begin{equation}
\langle|Z_{j\pm}|^2\rangle_\phi = 2\sigma_j^2\int_0^{2\tau}dt_1\int_0^{t_1}dt_2 s(t_1) s(t_2) F_{j\pm}(t_1-t_2),
\end{equation}
and the functions $F_{j\pm}(t)$ are:
\begin{equation}
F_{j\pm}(t) = e^{-t/T_\phi}\cosh\left[\left(i\omega_{j\pm}+2/T_\phi\right)t\right].
\end{equation}

The remaining coefficients in the matrix representation of $\left<\mathcal{L}_0\right>_\phi$, $\left<\mathrm{Re} Z^2\right>_\phi$ and $\langle\mathrm{Im} Z^2\rangle_\phi$, are given by
\begin{eqnarray}
\langle\mathrm{Re} Z^2\rangle_\phi & = & \sum_j \sigma_j^2\int_0^{2\tau}dt_1\int_0^{t_1}dt_2 K_j^+(t_1,t_2),\\
\langle\mathrm{Im} Z^2\rangle_\phi & = & -i\sum_j \sigma_j^2\int_0^{2\tau}dt_1\int_0^{t_1}dt_2 K_j^-(t_1,t_2),
\end{eqnarray}
with integral kernels
\begin{multline}
K_j^\pm(t_1,t_2) = s(t_1)s(t_2)\cos\left[\omega_j(t_1-t_2)\right]e^{(t_1-t_2)/T_\phi}\times\\
\times\left[e^{i\gamma_H B(t_1+t_2)}\pm e^{-i\gamma_H B(t_1+t_2)-4(t_1+t_2)/T_\phi}\right].
\end{multline}
The integrals can all be evaluated analytically, but we leave them unevaluated here for notational convenience.

\section{Filter functions for dynamical decoupling}\label{sec:DDFilterFunctions}
For an $n$-pulse PDD-X sequence, we find the generalized filter function from direct integration of Eq.~\eqref{eq:TwoAxisFilter}:
\begin{multline}\label{eq:PDDXFilter}
Z_{j\pm}[(n+1)\tau] = \frac{\sigma_j}{\omega_{j\pm}}\tan\left(\frac{\omega_{j\pm}\tau}{2}\right)\times\\
\times\left[1+(-1)^n e^{i\omega_{j\pm}(n+1)\tau}\right].
\end{multline}
Here, $\omega_{j\pm}=\omega\pm\omega_j$, where $\omega=\gamma_H B$ gives the hole-spin Zeeman splitting and $\omega_j=\gamma_j B$ determines the Zeeman splitting of nuclear-spin species $j$.

For an $n$-pulse PDD-Y sequence, integrating Eq.~\eqref{eq:TwoAxisFilter} gives
\begin{equation}\label{eq:PDDYFilter}
Z_{j\pm}[(n+1)\tau]=\frac{2\sigma_j}{\sin\omega_j \tau}e^{i\omega\tau/2}e^{\pm i\omega_j(n+1)\tau/2} G_n^\pm(\tau),
\end{equation}
where
\begin{multline}
G_n^\pm(\tau) = \frac{\tau}{2}\left[\sin\left(\frac{\omega_j(n+2)\tau}{2}\right)\mathrm{sinc}\left(\frac{\omega_{j\pm}\tau}{2}\right)-\right.\\
-\left.\sin\left(\frac{\omega_j n\tau}{2}\right)\mathrm{sinc}\left(\frac{\omega_{j\mp}\tau}{2}\right)\right],\quad[n\,\mathrm{even}], 
\end{multline}
and
\begin{multline}
G_n^\pm(\tau) = \frac{\tau}{2}\sin\left(\frac{\omega_j(n+1)\tau}{2}\right)\left[e^{\mp i\omega_j\tau/2}\mathrm{sinc}\left(\frac{\omega_{j\pm}\tau}{2}\right)-\right.\\
-\left.e^{\pm i\omega_j\tau/2}\mathrm{sinc}\left(\frac{\omega_{j\mp}\tau}{2}\right)\right],\quad[n\,\mathrm{odd}]. 
\end{multline}

\bibliography{purity-anisotropic}
\end{document}